\documentclass
[nofootinbib,superscriptaddress,aps,prd,showkeys,noshowpacs,onecolumn,10pt]{revtex4-2}%
\usepackage{graphics}
\usepackage{graphicx}
\usepackage{subcaption}
\usepackage{epsf}
\usepackage{bm}
\usepackage{amsmath,amssymb,amsfonts,mathrsfs,amsthm}
\usepackage{latexsym}
\usepackage{enumerate}
\usepackage{comment}
\usepackage[dvipsnames,svgnames,x11names,table]{xcolor}
\usepackage[colorlinks = true,
            linkcolor = Cerulean,
            urlcolor  = magenta,
            citecolor = magenta,
            anchorcolor = NavyBlue]{hyperref}
\usepackage{epstopdf}%
\usepackage{bbm}

\def\be{\begin{equation}}
\def\ee{\end{equation}}

\newcommand{\ima}{\mathrm{i}}
\makeatother

\begin{document}

\title{Dirac-Bergmann algorithm and canonical quantization of $k$-essence cosmology}
\author{Andr\'es Lueiza-Colip\'i}
\email{a.lueiza01@ufromail.cl}
\affiliation{Departamento de Ciencias F\'{\i}sicas, Universidad de La Frontera, Casilla
54-D, 4811186 Temuco, Chile}
\author{Andronikos Paliathanasis}
\email{anpaliat@phys.uoa.gr}
\affiliation{Institute of Systems Science, Durban University of
Technology, Durban 4000, South Africa}
\affiliation{Centre for Space Research, North-West University, Potchefstroom 2520, South Africa}
\affiliation{National Institute for Theoretical and Computational Sciences (NITheCS), South Africa}
\affiliation{Departamento de Matem\'{a}ticas, Universidad Cat\'{o}lica del Norte, Avda.
Angamos 0610, Casilla 1280 Antofagasta, Chile}
\author{Nikolaos Dimakis}
\email{nikolaos.dimakis@ufrontera.cl}
\affiliation{Departamento de Ciencias F\'{\i}sicas, Universidad de La Frontera, Casilla
54-D, 4811186 Temuco, Chile}

\begin{abstract}\noindent
   We develop a general canonical quantization scheme for $k$-essence cosmology in scalar-tensor theory. Utilizing the Dirac-Bergmann algorithm, we construct the Hamiltonian associated with the cosmological field equations and identify the first- and second-class constraints. The introduction of appropriate canonically conjugate variables with respect to Dirac brackets, allows for the canonical quantization of the model. In these new variables, the Hamiltonian constraint reduces to a quadratic function with no potential term. Its quantum realization leads to a Wheeler-DeWitt equation reminiscent of the massless Klein-Gordon case. As an illustrative example, we consider the action of a tachyonic field and investigate the conditions under which a phantom crossing can occur as a quantum tunneling effect. For the simplified constant potential case, we investigate the consequences of different boundary conditions on the singularity avoidance and to the mean expansion rate.
\end{abstract}

\maketitle

\section{Introduction}

In the early stages of the universe, when the spacetime curvature becomes extremely large and the associated curvature radius correspondingly small, quantum phenomena are expected to play a significant role in the structure of the cosmos \cite{Fischetti:1979ue,Hartle:1979uf,Hartle:1980nn,Hartle:1980kz}. In this regime, the classical description of gravity based on general relativity breaks down, implying the need for a quantum theory of gravity. Although a complete and universally accepted framework for such a theory is still elusive, quantum cosmology is formulated as a viable approximation for studying early cosmological phenomena \cite{Kiefer:2025udf}. Some of the various proposals in the literature for the quantization of the gravitational field include canonical quantization \cite{DeWitt1967,Kuchar}, path integrals \cite{Hawking:1978jz,Teitelboim:1981ua}, loop models \cite{Ashtekar:2008zu,Rovelli}, spin-foams \cite{Perez:2012wv,Asante} and others \cite{Vilenkin:1994rn,Bojowald:2015iga,Kiefer:2008sw,Reuter,Surya,Calcagni}.

The fundamental dynamical object in quantum cosmology is the \textit{wavefunction of the universe} \cite{Hartle:1983ai}, which is intended to encode information about the possible quantum states of the early Universe. However, the physical interpretation of this wave function remains uncertain and is still the subject of an ongoing debate. Despite these conceptual challenges, quantum cosmology provides a useful framework to investigate fundamental problems of early-Universe physics, such as the onset of cosmic inflation, the initial-value problem, and the Big Bang singularity \cite{Misner,Berger,Lemos,DeWitt1967,Bojowald:2015iga,Vilenkin2002,Vilenkin1988,Halliwell1990,Halliwell1988,Linde1990,Handley2021,Kiefer:2010zzb,
Kiefer:2022,Vakili,Zampeli:2015ojr,QC1,QC2,QC3,QC4,QC5,Matone,Dimakistime,QCextra1}. Furthermore, within the framework of Bohmian representation of quantum theory \cite{Bohm:1951xw,Bohm:1951xx}, the wavefunction has been used for the derivation of quantum corrections to the classical gravitational solution in the semi-classical limit \cite{
Pinto-Neto:2018zvn,Vicente:2021abv,Paliathanasis:2017ocj,Paliathanasis:2018ixu,Pinto-Neto:2021gcl,Vicente:2023hba,Basilakos:2025hyk}. 

Cosmic inflation is a mechanism where the universe expands dramatically in the very early stages of the universe, this scenario was originally proposed to address fundamental shortcomings of the standard Big Bang cosmology, most notably the horizon and flatness problems \cite{Starobinsky1980,Guth1981,Linde1982,AlbrechtSteinhardt1982}. The inflationary phase eventually ends through a reheating process, during which the energy stored in the inflationary sector is transferred to standard matter fields, giving rise to a hot, radiation dominated universe \cite{AlbrechtReheating1982,Kofman1994,Kofman1997}. Cosmic inflation is not predicted by the standard cosmological $\Lambda$CDM model, an early dominating universe component is needed to drive this accelerated expansion \cite{Barrow:1995xb,Lidsey:1991zp,Ford:1989me,Heusler:1991ep,Kim:2010fq,Lyth:1998xn,Adams:1997de,Thomas:1995dq,Binetruy:1986ss,Herrera:2024rcm,Luongo:2024opv,Luongo:2023aaq,DAgostino:2022fcx,Barrow:2016qkh}. Since inflation operates at very early times, it is often the object of study within the framework of quantum cosmology  \cite{Linde:1983mx,Linde:1995ck,Khoury:2022ish,Jalalzadeh:2022dlj}. In this setting, it is investigated the onset of the inflationary phase, its probability of occurrence, and the robustness of inflationary background solutions with respect to different choices of initial conditions \cite{Vilenkin2002,Vilenkin1988,Halliwell1990,Halliwell1988,Linde1990,Handley2021}.

The most widely studied class of inflationary models is single-field slow-roll inflation, in which the accelerated expansion of the universe is driven by a scalar field (usually named the inflaton) whose energy density is dominated by its potential \cite{Guth1981,Linde1982,Martin:2013tda}. During inflation, the inflaton slowly rolls toward a minimum of its potential, leading to a graceful exit from the inflationary phase. This class of models remains in good agreement with current observations of the cosmic microwave background (CMB) \cite{Balkenhol2025}. Nevertheless, the search for alternative inflationary scenarios remains an active area of research. One such alternative, originally proposed in \cite{ArmendarizPicon1999} is the \textit{$k$-inflation}, in which the accelerated expansion is driven by a scalar field with non-canonical kinetic terms, without the need for a potential. The motivation for considering non-canonical kinetic terms arises more or less naturally in high-energy theories, particularly in string theory from the Dirac-Born-Infeld action \cite{ASen1999}.

Recent cosmological results from the Dark Energy Spectroscopic Instrument (DESI) constrain the late-time evolution of dark energy and show a mild preference for departures from a cosmological constant, including a possible phantom behaviour for dark energy \cite{DESI2025_DR2}. The theoretical mechanism that describes the observational data is still unknown, but there are various proposals in the literature, which consider quantum effects \cite{Li:2025cxn,Oriti:2021rvm,Paliathanasis:2025dcr,Paliathanasis:2025kmg,Singh:2019hhi} for the description of dark energy. In \cite{Dimakis:2020tzc} it has been shown that, in the case of a multi-scalar field cosmology, the equation-of-state parameter for the cosmic fluid can cross the phantom divide line as an effect of quantum transitions.

$K$-essence theories can also account for the present accelerated expansion of the universe, through a dynamical dark energy that naturally drives the accelerated expansion \cite{ArmendarizPicon2001}, because they allow the unification of the dark sector of the universe. That is, $k$-essence theories can describe dark matter and dark energy, and also unify the inflationary epoch with the late-time acceleration \cite{Bose:2008ew,Bose:2009kc,Saa}. Recently, in \cite{Csillag:2025gnz} it has been demonstrated that $k$-essence theories with quadratic terms arise naturally in extended theories of gravity, when the connection that describes the gravitational field is not the Levi-Civita connection. The tachyonic field constitutes a special class of $k$-essence theory \cite{Bagla:2002yn}. The loop quantum cosmology of the tachyonic field was investigated in \cite{Xiong:2007ak}, while loop quantum $k$-essence models are presented in \cite{Shi:2021dpi}.

In this work we examine the canonical quantization of $k$-essence. To realize this analysis we use the Dirac–Bergmann algorithm and identify all the first- and second-class constraints present in the Hamiltonian description. The algorithm has been applied before in extended theories of gravity, see for instance \cite{Ferraro:2018tpu,Tomonari:2023wcs,Dimakis:2021gby,Dimakis:2023oje,DAmbrosio:2023asf,Tomonari:2024ybs} and references therein. We show that the Hamiltonian constraint, in terms of the new appropriate canonically conjugate variables, is expressed as a purely quadratic function with no potential term. This leads to a quantum Wheeler-DeWitt constraint, which is formulated as a massless Klein-Gordon (KG) equation in $1+1$ dimensions. The case of a tachyonic field with a constant potential is examined in detail, where we explicitly derive the wave function of the universe. Surprisingly, the quantization process is extremely well-behaved in the sense of producing a finite probability, which is rare occurrence in the canonical study of cosmological configurations. Through the analysis of the probability amplitudes of the tachyonic field we investigate whether phantom crossing can emerge due to quantum effects. Furthermore, the avoidance of singularities is discussed as well as the consequence of different choices of boundary conditions. The structure of the paper is as follows.

In Section II we present the basic elements and definitions of the $k$-essence theory. Its mini-superspace description within an isotropic and homogeneous cosmological background is given in Section III. Furthermore, the Dirac–Bergmann algorithm is introduced in Section IV. We employ the algorithm and derive a quadratic Hamiltonian function, which allows us to define the Wheeler-DeWitt equation. In Section V we discuss the canonical quantization for the tachyonic field with a constant potential function. We derive the analytic solution and the wave function given by the resulting Wheeler-DeWitt equation. Finally, in Section VI we draw our conclusions.

\section{Theory and Field Equations}

In this section we introduce the basic concepts of the $k$-essence theory and the corresponding field equations; for further details we refer to \cite{ArmendarizPicon2001}. The action integral representing a generic $k$-essence theory is of the form 
\begin{equation}
  S = \int\!\! \sqrt{-g} F(R,X,\phi) d^4 x \, ,
\end{equation}
where $R$ is the Ricci scalar curvature, $\phi$ the scalar field and $X$ represent its kinetic term defined as
\begin{equation}\label{kin}
  X=-\frac{1}{2} \nabla_\mu \phi \nabla^\mu \phi \, .
\end{equation}
Variation of the action with respect to the metric results in the subsequent gravitational field equations \cite{Bahamonde2015}
\begin{equation} \label{feqmet}
   F_R \left(R_{\mu\nu} - \frac{1}{2} g_{\mu\nu}R\right) - \left[ \frac{1}{2} g_{\mu\nu} \left( F - R F_R \right) + \nabla_\mu \nabla_\nu \left( F_R \right) - g_{\mu\nu} \nabla_\kappa\nabla^\kappa \left( F_R \right)  \right] - \frac{1}{2} F_X \nabla_\mu \phi \nabla_\nu \phi =0 \, ,
\end{equation}
while variating  with respect to the scalar field $\phi$ yields
\begin{equation} \label{feqphi}
   \nabla_\kappa \left( F_X \nabla^\kappa \phi \right) + F_\phi =0 \, .
\end{equation}
In the above relations, subscripts $F_R$, $F_X$ and $F_\phi$ denote partial derivatives of $F$ with respect to the indicated variables. 

In this work, we focus on a general class of theories of the form 
\begin{equation}
  F(R,X,\phi) = f_1(\phi) R+ f_2(X,\phi) \, .
\end{equation}
That is, theories involving a linear term in the scalar curvature, which however is not necessarily minimally coupled to the scalar field. Furthermore, the kinetic term of the scalar field is meant to be included through non-linear terms in the action, in order to distinguish our analysis from that of a standard canonical scalar field theory. 

\section{Mini-superspace Lagrangian}

We assume a spatially flat Friedmann–Lemaître–Robertson–Walker spacetime with the line element
\begin{equation}\label{lineel}
  ds^2 = -N (t)^2 dt^2 + a(t)^2 \left(dr^2 + r^2 d\theta^2 + r^2 \sin^2\theta d\varphi^2\right) \, ,
\end{equation}
and construct the corresponding mini-superspace Lagrangian by isolating the dynamical part of the resulting Lagrangian density
\begin{equation}
  \mathcal{L} \sim \sqrt{-g} \left( f_1(\phi) R+ f_2(X,\phi) \right) - \lambda \left( X + \frac{1}{2} \nabla_\mu \phi \nabla^\mu \phi \right) \, .
\end{equation}
We choose to treat $X$ as an independent field, which is linked to expression \eqref{kin} with the help of a Lagrange multiplier. This is crucial for performing the transition to the Hamiltonian description when the Lagrangian contains a non-linear dependence on $X$. 

After removing the acceleration terms $\ddot{a}$ by adding a total time derivative, and fixing the value of the Lagrange multiplier by calculating the equation of motion for $X$, we arrive at the point Lagrangian
\begin{equation} \label{Lag}
  L = \frac{1}{2N}\left( a^3  f_{2,X} \dot{\phi}^2 -12 a^2 f_1'(\phi)  \dot{a} \dot{\phi} -12 a f_1(\phi) \dot{a}^2  \right) +N a^3 \left(f_2(\phi,X)-X f_{2,X}\right) \, .
\end{equation}
The term $f_{2,X}$ stands for the partial derivative of $f_2(\phi,X)$ with respect to $X$ and the prime over $f_1(\phi)$ denotes the derivative of the latter with respect to its single argument. It is a matter of a straightforward calculation to demonstrate that the Euler-Lagrange equations of \eqref{Lag} are equivalent to the set of the field Eqs. \eqref{feqmet} and \eqref{feqphi}, together with the condition \eqref{kin}, thus providing a valid mini-superspace description for the original gravitational system. 

The set of Euler-Lagrange equations with respect to $N$, $a$, $\phi$ and $X$ reduces to 
\begin{align}
  & \frac{12}{N} \left( \frac{\dot{a} \dot{\phi} f_1^{\prime}}{a}+\frac{\dot{a}^2 f_1}{a^2} \right) -\frac{\dot{\phi}^2 f_{2,X}}{N} + 2 N \left(f_2-X f_{2,X}\right) =0 \, , \\
  & 8 f_1 \left(\frac{\ddot{a}}{a}-\frac{\dot{a} \dot{N}}{a N}\right) + 4 \dot{\phi} \left(\frac{2 \dot{a}}{a}-\frac{\dot{N}}{N}\right) f_1^{\prime} + \dot{\phi}^2 \left(f_{2,X}+4 f_{1}^{\prime\prime}\right)+\frac{4 \dot{a}^2 }{a^2}f_1 +4 \ddot{\phi} f_1^{\prime} \nonumber \\
  & -2 N^2 X f_{2,X}+2 N^2 f_2 =0 \, ,  \\
  & 6 \ddot{a} f_1^{\prime}-a \ddot{\phi} f_{2,X} + 6 \dot{a} \left(\frac{\dot{a}}{a}-\frac{\dot{N}}{N}\right) f_1^{\prime} + a \dot{\phi} \left(\frac{\dot{N}}{N}-\frac{3 \dot{a}}{a}\right) f_{2,X} - a \dot{\phi} \left(\dot{X} f_{2,XX}+\frac{1}{2} \dot{\phi} f_{2,X\phi}\right) \nonumber \\
  & + a N^2 \left( f_{2,\phi}-X f_{2,X\phi}\right) =0 \, ,\\ \label{kincon}
  & f_{2,XX} \left(2 N^2 X-\dot{\phi}^2\right) =0 \, ,
\end{align}
where $f_{2,XX}$ stands for the second derivative with respect to $X$ and $f_{2,X\phi}$ for the mixed. From the last equation of the set \eqref{kincon}, it is evident that as long as the theory is not linear in $X$, the latter is fixed to assume the expression corresponding to Eq. \eqref{kin}. It is thus crucial for the validity of this construction that $f_{2}$ be non-linear in $X$.

\section{Dirac-Bergmann algorithm and the Hamiltonian formalism} \label{DBgen}

From the form of the Lagrangian \eqref{Lag}, we distinguish two primary constraints. Due to the absence of ``velocities'' for $N$ and $X$, the corresponding momenta are zero
\begin{align}
   p_N & := \frac{\partial L}{\partial \dot{N}} =0 \Rightarrow p_N \approx 0 \, , \\
   p_X & := \frac{\partial L}{\partial \dot{X}} =0 \Rightarrow p_X \approx 0 \, ,
\end{align}
where the use of the symbol ``$\approx$'' is meant to indicate a weak equality in the Dirac sense. That is, the fact that we set the constraints equal to zero only when appearing outside of Poisson brackets. 

The rest of the momenta are 
\begin{equation}
  p_a:=\frac{\partial L}{\partial \dot{a}} \quad \text{and} \quad   p_\phi:=\frac{\partial L}{\partial \dot{\phi}}\, ,
\end{equation}
with the canonical Hamiltonian written as
\begin{equation} \label{Hcan}
  H_C = \dot{a}p_a + \dot{\phi} p_\phi -L = N \mathcal{H} \, ,
\end{equation}
where
\begin{equation}
  \mathcal{H} = \frac{1}{2 a \left(f_1 f_{2,X}+3 f_1^{\prime\; 2}\right)} \left(\frac{f_1}{a^2} p_\phi^2  -\frac{ f_{2,X}}{12} p_a^2-\frac{f_1^{\prime}}{a}p_a p_\phi \right) + a^3 \left(X f_{2,X}-f_2\right) \, .
\end{equation}
Recall that the functional dependencies are:  $f_1=f_1(\phi)$ and $f_2=f_2(\phi,X)$.  Note that the special case $f_1 f_{2,X}+3 f_1^{\prime\; 2}=0$, which leads to the vanishing of a denominator in the above expression, is excluded from our analysis since it corresponds to a linear in $X$ theory.

Following the Dirac-Bergmann algorithm \cite{Dirac,Bergmann}, the primary Hamiltonian is obtained with the addition of the two primary constraints together with the inclusion of two multiplier functions $u_N(t)$ and $u_X(t)$,
\begin{equation}\label{Hprimary}
  H_P = N \mathcal{H} + u_N p_N + u_X p_X \, .
\end{equation}
The introduction of $X$ with a Lagrange multiplier in the Lagrangian \eqref{Lag} leads to the inclusion of an additional constraint,  namely $p_X\approx 0$. This is different from the standard analysis of cosmological Lagrangians with canonical scalar fields, where only $p_N\approx 0$ appears as a primary constraint. 

At this point, according to the Dirac-Bergmann algorithm, we need to impose the consistency conditions requiring that the constraints are being preserved in time. This leads to the supplementary relations
\begin{equation}
  \dot{p}_N\approx 0\, , \quad \quad \dot{p}_X\approx 0 ,
\end{equation}
which may yield one of the following outcomes: a) being satisfied identically, b) fixing some of the multipliers in \eqref{Hprimary}, or c) generating additional constraints. The algorithm closes when no new constraints are produced.

The first condition is universal in minisuperspace cosmology and it implies the relation
\begin{equation}
  \dot{p}_N \approx 0 \Rightarrow \{p_N,H_P\} \approx 0 \Rightarrow \mathcal{H} \approx 0\, ,
\end{equation}
known as the Hamiltonian or quadratic constraint. We thus obtain $\mathcal{H}\approx 0$ as a secondary constraint. In addition to the latter, the second consistency condition yields
\begin{equation}
   \dot{p}_X \approx 0 \Rightarrow \{p_X,H_P\} \approx 0 \Rightarrow \chi \approx 0 \, .
\end{equation}
This is a new, quadratic in the momentum $p_a$, secondary constraint which reads
\begin{equation}\label{secchi}
  \chi := p_a^2 + 24 a^4 \left[ f_1 \left(f_2-2 X f_{2,X}\right)-3 X f_1^{\prime\; 2} \right] \approx 0 \, .
\end{equation}
The preservation in time must be invoked for the secondary constraints as well. Specifically, we require $\dot{\mathcal{H}} \approx 0$ and $\dot{\chi} \approx 0$. The first condition, for $\mathcal{H}$, is trivially satisfied since $\dot{\mathcal{H}}=\{\mathcal{H},H_P\}=0$. In the case of $\chi\approx 0$, the condition $\dot{\chi}\approx 0$ leads to the determination of the multiplier $u_X$ and thus the algorithm closes since no new constraint was generated. The analysis revealed the existence of four constraints: $p_N\approx 0$, $p_X \approx 0$, $\mathcal{H}\approx 0$ and $\chi\approx 0$.

For the quantization it is imperative to distinguish between first-class and second-class constraints. First-class constraints commute, at least weakly, with all other constraints and generally indicate the existence of a gauge freedom in the system. According to Dirac's prescription for the quantization of constrained systems, these are implemented as operators that annihilate the wave function. Conversely, second-class constraints have at least one non-zero bracket with other constraints and correspond to superfluous degrees of freedom that must be eliminated prior to quantization. A direct calculation shows that three of the constraints (all save $p_N\approx 0$) have non-vanishing\footnote{Note that non-vanishing here means even after imposing the constraints to be zero at the end result of the bracket.} Poisson brackets when taken in pairs. However, as is known from the theory \cite{Diracbook}, second class constraints appear only in even numbers. This implies that there exists a first class linear combination within the set of the three aforementioned constraints.  

We identify the linear combination of constraints
\begin{equation}\label{newcon}
  \begin{split}
  \bar{\mathcal{H}} := & 2 a^3 \left(f_1 f_{2,X}+3 f_1^{\prime\; 2}\right) \left(f_1 \left(f_{2,X}+2 X f_{2,XX}\right)+3 f_1^{\prime\; 2}\right) \mathcal{H} \\
  & + \Big[ 2 \left(\left(f_1^2 \left(f_{2,\phi}-2 X f_{2,X\phi}\right)+9 X f_1^{\prime\; 3}\right)-2 f_1 f_1^{\prime} \left(X \left(3 f_1^{\prime\prime}-2 f_{2,X}\right)+f_2\right)\right) \Big] p_X \approx 0 \, ,
  \end{split}
\end{equation}
as a first class relation. That is, a constraint that commutes (at least weakly) with all of the others. As a result, we arrive at the following classification: two first class constraints $p_N\approx 0$ and $\bar{\mathcal{H}} \approx 0$ (defined by \eqref{newcon}), and two second class $p_X\approx0$ and $\chi\approx0$. 

In order to proceed with the canonical quantization, and remove the superfluous degrees of freedom, we introduce the Dirac brackets. The latter allow us to set the second class constraints strongly equal to zero, effectively eliminating them from the system \cite{Diracbook,Sund}. 

First, we define the matrix whose elements are formed by the Poisson brackets of the second class constraints $p_X\approx0$ and $\chi\approx0$,
\begin{equation}
  \Delta = \begin{pmatrix}
             0 & \{p_X,\chi\} \\
             \{\chi,p_X\} & 0 
           \end{pmatrix} \, ,
\end{equation}
which is by construction invertible. The Dirac brackets, for two phase space quantities $A$, $B$ are defined as \cite{Diracbook,Sund} 
\begin{equation}\label{dbra}
  \{A,B\}_D = \{A,B\} - \{A,(\text{s.c.})_{i}\} \Delta^{-1}_{\; ij} \{(\text{s.c.})_{j}, B\} \, ,
\end{equation}
where $(\text{s.c.})_{i}$ denotes the second class constraints, namely $(\text{s.c.})_{1}=p_X$ and $(\text{s.c.})_{2}=\chi$. The Dirac bracket effectively converts the second class constraints into strong equations because every phase space function commutes with them under this bracket. 

We can now reduce the phase space and eliminate the second class constraints by substituting them as strong equations in the Hamiltonian. The initial phase space was spanned by $(N,a,\phi,X)$ and their respective momenta. By using $p_X=0$ and $\chi=0$ as strong equations, we eliminate two of the eight variables and the dimensionality is reduced by the same amount. The condition $p_X=0$ obviously eliminates the momentum for $X$. The $\chi$, as seen from Eq. \eqref{secchi}, can be used to remove $p_a$ since it has a simple quadratic dependence with respect to the latter. We are thus left with $(a,\phi,X,p_\phi)$ and of course $(N,p_N)$. The former however do not constitute a set of canonically conjugate variables with respect to the Dirac brackets. To amend this, we introduce the variables
\begin{subequations} \label{newvar}
\begin{align} 
  \pi_X & = \mp \sqrt{\frac{2}{3}} \frac{ a^3 \left(f_1 \left(f_{2,X}+2 X f_{2,XX}\right)+3 f_1^{\prime\; 2}\right)}{\left[3 X f_1^{\prime\; 2}-f_1 \left(f_2-2 X f_{2,X}\right)\right]^{\frac{1}{2}}} \, ,\\
  \pi_\phi & = p_\phi \pm \sqrt{\frac{2}{3}}\frac{ a^3 \left(f_1^{\prime} f_2-2 X f_1^{\prime} \left(f_{2,X}+3 f_1^{\prime\prime}\right)+f_1 \left(f_{2,\phi}-2 X f_{2,X\phi}\right)\right)}{\left[3 X f_1^{\prime\; 2}-f_1 \left(f_2-2 X f_{2,X}\right)\right]^{\frac{1}{2}}} \, ,
\end{align}
\end{subequations}
which satisfy
\begin{equation}
 \begin{split}
    & \{ X, \pi_X \}_D=1=\{ \phi, \pi_\phi \}_D \, ,\\
    & \{ X, \phi \}_D=\{ X, \pi_\phi \}_D = \{ \phi, \pi_X \}_D = \{ \pi_X, \pi_\phi \}_D =0 \, .
   \end{split}
\end{equation}
The plus/minus sign in Eqs. \eqref{newvar} depends on the sign of the square root used in the substitution of $p_a$ from $\chi=0$. So, apart from $(N,p_N)$ we adopt as basic phase space variables $X$, $\phi$ and their canonical conjugates $\pi_X$ and $\pi_\phi$ in place of $a$ and $p_\phi$.

The resulting Hamiltonian constraint \eqref{newcon} in this reduced phase space takes the form
\begin{equation} \label{hamred}
  \mathcal{H}_{red}= G^{ij}\pi_i \pi_j \approx 0, \quad\quad i,j=1,2 \, ,
\end{equation}
where $\pi_i=(\pi_X, \pi_\phi)$ and the components of the induced inverse ``mini-supermetric'' are 
\begin{equation}
  \begin{split}
     G^{11} = & \frac{1}{\left(f_{2,X}+2 X f_{2,XX}\right) f_1  +3 f_1^{\prime\; 2}} \Bigg[3 X  \left(f_2-2 X f_{2,X}\right) f_1^2 f_{2,X}^2 \\
     & -9 X  \left(4 X \left(f_{2,X}+3 f_1^{\prime\prime}\right)+f_2\right) f_1^{\prime \; 4}  \\
     & +  \left(2 X f_{2,X} \left(f_2-24 X f_1^{\prime\prime}\right)+4 \left(f_2+3 X f_1^{\prime\prime}\right)^2-29 X^2 f_{2,X}^2\right) f_1 f_1^{\prime\; 2} \\
     & +\left(f_{2,\phi}-2 X f_{2,X\phi}\right) \Big( \left(9 X f_1^{\prime\; 2}-2 f_1 \left(-2 X f_{2,X}+f_2 +3 X f_1^{\prime\prime}\right)\right) 2 f_1 f_1^{\prime} \\
     & + \left(f_{2,\phi}-2 X f_{2,X\phi}\right) f_1^3 \Big) \Bigg] \, ,
   \end{split} 
\end{equation}
\begin{equation}
  G^{12} = -2  \left(-2 X f_{2,X}+f_2+3 X f_1^{\prime\prime}\right)f_1 f_1^{\prime} + \left(f_{2,\phi}-2 X f_{2,X\phi}\right)f_1^2+9 X f_1^{\prime \; 3} \, ,
\end{equation}
\begin{equation}
  G^{22} = \left(\left(f_{2,X}+2 X f_{2,XX}\right)f_1 +3 f_1^{\prime\; 2}\right) f_1 \, .
\end{equation}
Even though the final expression is rather involved, it is quite remarkable that the final constraint consists of just a ``kinetic'' term. The consequence of this is that the corresponding Wheeler-DeWitt equation, obtained by choosing the Laplacian as our basic operator, has the form of a massless Klein-Gordon equation
\begin{equation}\label{WDW}
  \widehat{\mathcal{H}}\psi = \frac{-\hbar^2}{\sqrt{|G|}} \partial_i \left( \sqrt{|G|}\, G^{ij} \partial_j \Psi \right) =0 \, .
\end{equation}
The $G$ denoting the determinant of the mini-supermetric $G=\mathrm{det}(G_{ij})$, with the indices $i$, $j$ running in the two coordinates $X$ and $\phi$. The space spanned by $G_{ij}$ in $X$ and $\phi$ is of course conformally flat since is two-dimensional.

With regards to the aforementioned conformally flatness of the configuration space, we can make an interesting observation. There of course exist coordinates, say $(\phi,X)\mapsto(u,v)$, where the mini-superspace metric appearing in \eqref{hamred} assumes the form
\begin{equation}
  G_{ij} = g(u,v) \eta_{ij} \, ,
\end{equation}
where $\eta_{ij}=\text{diag}(-1,1)$. Recall that in the construction of the first class combination \eqref{newcon}, from which $\mathcal{H}_{red}$ is later obtained, we have the freedom of using an overall multiplying factor without changing the first class nature of the constraint. Consequently, it is possible to construct a first class combination such that the function $g(u,v)=1$. The resulting reduced Hamiltonian constraint in the new variables is then just
\begin{equation} \label{hamredflat}
  \mathcal{H}_{red}= -\pi_u^2 + \pi_v^2 \approx 0, \quad\quad i,j=1,2 \, ,
\end{equation}
leading to a quantization in a flat 2d-plane. The distinction among distinct $k$-essence theories thus arises from the different mappings $(\phi,X)\mapsto(u,v)$, and their domains of definition. This is expected to allow different interpretations of the results even though the expression of the wave function in the $u,v$ variables is the same for all theories. In the next section, we illustrate this process using the tachyon case.

\section{The cosmic tachyon}

\subsection{Classic description}

In this section, we consider the rolling tachyon, which was introduced and studied in \cite{Sen1,Sen2,Gibbons,inflt1}. The function $F(R,X,\phi)$ characterizing the Lagrangian density of the theory is
\begin{equation}
   F(R,X,\phi) = \frac{R}{2\kappa} - V(\phi)\sqrt{1-2 X} \, .
\end{equation}
The mini-superspace Lagrangian, which correctly reproduces the field equations, is given by
\begin{equation}
  L = \frac{1}{2 N} \left(\frac{a^3 V(\phi)}{\sqrt{1-2 X}}\dot{\phi}^2 - \frac{6}{\kappa}a \dot{a}^2\right) - \frac{N  a^3 V(\phi)}{\sqrt{1-2X}} \left( 1- X \right) \, .
\end{equation}
It is straightforwardly derived from Eq. \eqref{Lag} by substituting $f_1=1/(2\kappa)$ and $f_2=- V(\phi)\sqrt{1-2 X}$. The resulting equations of motion are equivalent to the subsequent system:
\begin{subequations} \label{eqofmotach}
\begin{align} \label{eqofmotach1}
  & \left(\frac{\dot{a}}{N a}\right)^2 - \frac{\kappa }{3} \frac{ V(\phi)}{\sqrt{1-2X}} =0 \, ,  \\ \label{eqofmotach2}
  & \frac{\ddot{a}}{N^2 a}-\frac{\dot{N} \dot{a}}{N^3 a} - \frac{\kappa}{3}  V(\phi)\frac{ 1-3X }{\sqrt{1-2 X}} =0 \, ,\\ \label{eqofmotach3}
  & \frac{\dot{X}}{6 N X} + \left(1-2 X\right)\frac{\dot{a}}{N a} + \left(1-2 X\right) \frac{V'(\phi)}{3 \sqrt{2X} V(\phi)} =0 \, ,
\end{align}
\end{subequations}
where we have used $X=-\frac{1}{2} \nabla_\mu \phi \nabla^\mu \phi=\frac{\dot{\phi}^2}{2 N}$. The first two equations of the set \eqref{eqofmotach} correspond to the Friedmann equations, while the last is the equation of motion for the tachyon. They are expressed in a time parametrization invariant form since they still involve the lapse function $N$. 

The effective pressure and energy density of the tachyon are respectively
\begin{equation} \label{rhopeff}
  p = - V(\phi) \sqrt{1-2 X} , \quad \rho = \frac{V(\phi)}{\sqrt{1-2 X}},
\end{equation}
and the fluid equation of state parameter is given by
\begin{equation} \label{weff}
  w = \frac{p}{\rho} = -1 + 2 X .
\end{equation}
In inflationary scenarios involving tachyons \cite{inflt1}, the potential is considered constant near the region of unstable equilibrium, where $X\sim 0$, driving, as is evident from \eqref{weff}, an accelerating expansion ($w\simeq -1$).  

In what follows, we study the specific case 
\begin{equation}
  V(\phi)=V_0= \text{const.} \, ,
\end{equation}
with the means to solve analytically both the classical system, as well as the quantum Wheeler-DeWitt equation. It should be noted, of course, that this is not a realistic model by itself. However, as we previously mentioned, the limit $V(\phi)\simeq$const. is quite important in the slow-roll approximation of tachyonic inflation.

\begin{figure}[h]
  \centering
  \includegraphics[scale=0.7]{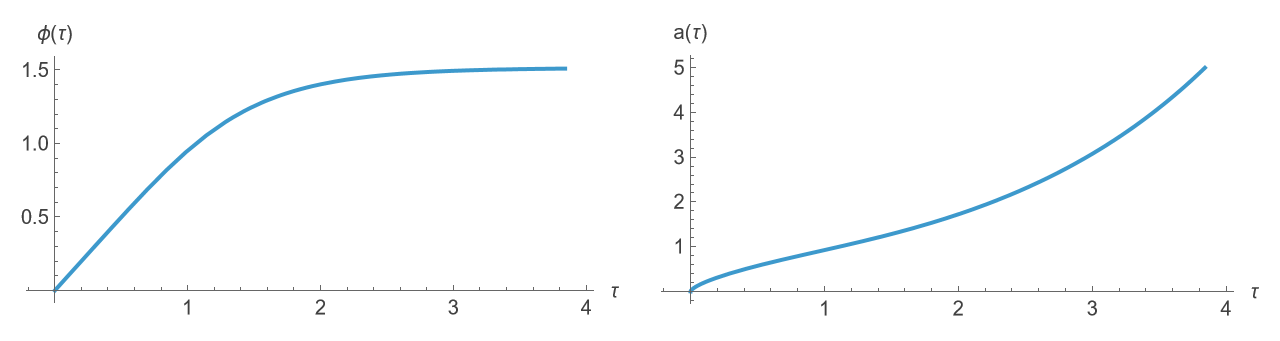}
  \caption{The plots of the tachyon field and the scalar factor as functions of the cosmic time. The value of $\phi$ is bounded.} \label{fig1}
\end{figure}

\begin{figure}[h]
  \centering
  \includegraphics[scale=0.7]{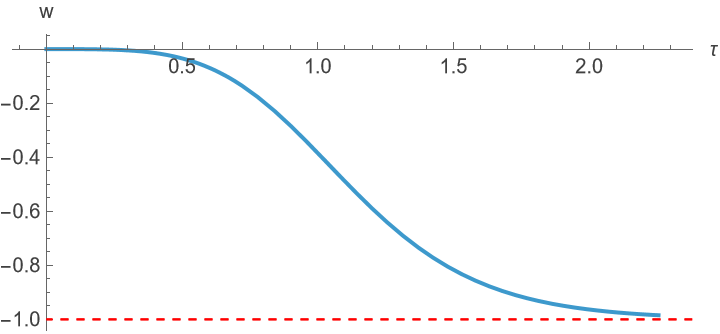}
  \caption{The effective equation of state parameter as a function of the cosmic time.} \label{fig2}
\end{figure}

It has been previously noted in the literature \cite{Gibbons2} that the $V=$const. case is equivalent to a Chaplygin gas description. Although the system is simple enough, the solution cannot be expressed in closed form in the cosmic time gauge $N=1$. We can surpass this difficulty and derive an analytic expression if, instead of the cosmic time, we adopt the scale factor as an effective time variable. That is, in place of fixing the time gauge by setting $N=1$, we choose the scale factor $a$ to be the time, i.e. set $a(t)=t$. The lapse function corresponding to this time gauge is then obtained from the field equations. In what follows, we present the resulting expressions which solve the system of Eqs. \eqref{eqofmotach}:
\begin{subequations} \label{classicsol}
\begin{align}
   & a(t) = t \, , \\
   & N(t) = \sqrt{\frac{3}{\kappa\, V_0 \, t^2}}\left(\frac{c_1 t^6}{c_1 t^6+1}\right)^{\frac{1}{4}}  \, , \\ \label{solclphi}
   & \phi(t) = \phi_0 + \frac{2 c_1^{\frac{1}{4}} t^{\frac{3}{2}} }{ \sqrt{3\, \kappa\, V_0}} \, _2F_1\left(\frac{1}{4},\frac{3}{4},\frac{5}{4};-c_1 t^6\right) \, .
\end{align}
\end{subequations}
The $c_1$ and $\phi_0$ are constants of integration, while the $_2F_1(a,b,c;z)$ stands for the Gauss hypergeometric function. The cosmic time, $\tau$, corresponding to the gauge $N=1$ can be recovered through the relation
\begin{equation} \label{cosmicgauge}
  \tau = \int\!\! N(t) dt .
\end{equation}
In principle, the above relation could be inverted to yield $t(\tau)$ and, through this, obtain $a(\tau)$, i.e. the scale factor as a function of the cosmic time (gauge $N=1$). Such an inversion, however, is not practically feasible due to \eqref{cosmicgauge} being transcendental. Nevertheless, we can still study the relevant dynamical quantities by using the corresponding parametric plots that involve $\tau(t)$. In Fig. \ref{fig1}, we observe the tachyon $\phi(t)$ and the scale factor $a(t)$ plotted with respect to the $\tau(t)$. The $w$ as function of $\tau$ can be encountered in Fig. \ref{fig2}. Evidently, the constant $V=V_0$ model leads to a late time accelerated expansion, as $w$ approaches the cosmological constant limit $w=-1$. In both figures, the numerical values used for the constants of integration are $\phi_0=0$, $V_0=\kappa^{-1}$ and $c_1=1$. We observe from the first plot of Fig. \ref{fig1} that the scalar field takes values within a finite range. The maximum value can be calculated analytically from the solution \eqref{solclphi}, and for $\phi_0=0$ is given by
\begin{equation}\label{maxphi}
   \phi_{max} = 2 \sqrt{\frac{\pi }{3\kappa V_0}} \frac{ \Gamma \left(5/4\right)}{ \Gamma \left(3/4\right)},
\end{equation}
where $\Gamma$ stands for the gamma function. 

Regarding the variable $X$, representing the kinetic energy $X=\dot{\phi}^2/(2N^2)$, we can derive from the solution \eqref{classicsol} that it varies monotonically from $1/2$, when $t=0=\tau$, to $0$ at $t,\tau\rightarrow +\infty$. At a first glance, we may argue that $X$ is also bounded, since in this case, it is restricted in the region $0 \leq X<1/2$. This behavior, however, holds strictly for positive values of the constant $c_1$ in solution \eqref{classicsol}.

The positivity of the constant $c_1$ guarantees a real and Lorentzian solution. Nevertheless, we notice that for negative values of $c_1$, we may still obtain a Lorentzian solution if $t>1/(-c_1)^{1/6}$. The scalar field in this case turns phantom as it becomes imaginary; still with its modulus remaining bounded. To be more precise, the scalar field of \eqref{solclphi} assumes complex values with a constant real part and the latter can be eliminated by making use of the integration constant $\phi_0$. Hence, without loss of generality, we may state that $\phi$ becomes a purely imaginary quantity. In this scenario, the resulting $X$ variable becomes negative and unbounded ($X<0$), leading to a phantom field behaviour which drives the equation of state parameter beyond the phantom line, $w<-1$. In table \ref{tab1}, we list these two possibilities. We designate the first case, $0<X<1/2$, corresponding to a canonical scalar, as the Right region (R) and the other, $X<0$, as the Left region (L), with respect to their positioning on the $X=0$ axis. We could interpret (L) as a classically ``forbidden'' region since one of the variables of the phase space variables, namely $\phi$, assumes (on mass-shell) imaginary values there. 

\begin{table}[h!]
\centering
\begin{tabular}{|c|c||c|c|c|c|}
  \hline
  Region & conditions & metric & $\phi$ & $X$ & $w$ \\ \hline\hline
  (R) & $c_1>0$ & Lorentzian & real & $0<X<1/2$ & $-1<w<0$ \\ \hline
  (L) & $c_1<0$ and $t>1/(-c_1)^{1/6}$ & Lorentzian & imaginary & $X<0$ & $w<-1$ \\ \hline
\end{tabular}
\caption{The behavior of the fundamental functions of solution \eqref{classicsol}.} \label{tab1}
\end{table}

\subsection{Quantum analysis}

Let us continue with the quantization of the model. For this, we need to follow the steps described previously in the Hamiltonian description of the general theory. The primary Hamiltonian is given by Eq. \eqref{Hprimary}, where the Hamiltonian constraint now reads
\begin{equation} \label{contach}
  \mathcal{H} = \frac{\sqrt{1-2 X} p_\phi^2 }{2 a^3 V(\phi )} - \frac{\kappa  p_a^2}{12 a} -\frac{a^3 (X-1) V(\phi )}{\sqrt{1-2 X}} \approx 0 \, . 
\end{equation}
We initially work with the relations involving a generic potential $V(\phi)$ and later, at the level of the quantum equations, we shall restrict again the problem to $V=V_0$. The other secondary constraint of the theory, Eq. \eqref{secchi}, is
\begin{equation}\label{seccontach}
   \chi = p_a^2 -\frac{12 a^4 V(\phi )}{\kappa  \sqrt{1-2 X}} \approx 0 \, .
\end{equation}
A linear first-class combination of constraints, similar to \eqref{newcon}, is obtained by introducing
\begin{equation}\label{newcontach}
   \bar{\mathcal{H}} = \frac{a^3 V(\phi)^2}{\sqrt{1-2 X}} \mathcal{H} +\sqrt{1-2 X} \left( \kappa  a  X V(\phi )^2p_a-\sqrt{1-2 X} V'(\phi)p_\phi\right) p_X \, .
\end{equation}
As commented at the end of section \ref{DBgen}, this first-class constraint is chosen so as to lead to a flat mini-superspace metric at the level of the reduced Hamiltonian. At this stage, the first-class constraints are $p_N\approx 0$ and $\bar{\mathcal{H}}\approx 0$, while the second-class $p_X \approx 0$ and $\chi\approx 0$. 

In order to perform the reduction of the Hamiltonian, we define the (Dirac bracket \eqref{dbra}) canonically conjugate variables (see also Eq. \eqref{newvar})
\begin{equation} \label{newvartach}
  \pi_X = \pm \frac{a^3 \sqrt{V(\phi )}}{\sqrt{3\kappa } (1-2 X)^{5/4}}, \quad  \pi_\phi = p_\phi \pm \frac{a^3 V'(\phi )}{\sqrt{3\kappa } \left(1-2 X\right)^{\frac{1}{2}} \sqrt{V(\phi )}} \, .
\end{equation}
The plus sign in the above expressions is compatible with the negative root of \eqref{seccontach}, while the minus with the positive. Both choices lead to the same reduced Hamiltonian
\begin{equation}\label{Hredtach}
 \begin{split}
  \mathcal{H}_{red} = & \left(\frac{(1-2 X)^2 V'(\phi )^2}{2V(\phi )}-3 \kappa  (1-2 X)^{3/2} X V(\phi )^2\right) \pi_X^2 \\
  & + (2 X-1) V'(\phi )\pi_X \pi_\phi + \frac{V(\phi )}{2} \pi_\phi^2 \approx 0 \, .
  \end{split}
\end{equation}
As described in the general theory, this is a purely quadratic function of the momenta \eqref{newvartach}. In the case  $V(\phi)=V_0$, the induced mini-superspace metric, which we read from Eq. \eqref{hamred}, corresponds to
\begin{equation} \label{ministach}
  G_{ij} = \begin{pmatrix}
             \frac{1}{3 \kappa  V_0^2 (1-2 X)^{3/2} X} & 0 \\
             0 & \frac{2}{V_0}
           \end{pmatrix}  \, ,
\end{equation}
and it is flat.

With the use of the Laplacian as the differential operator for the above quadratic expression, the resulting Wheeler-DeWitt equation (see Eq. \eqref{WDW}) is
\begin{equation}
  -6 \hbar^2 \kappa \, V_0 X (1-2 X)^{3/2} \partial_{X,X}\Psi +3 \hbar^2 \kappa \,  V_0 (5 X-1) \sqrt{1-2 X} \partial_X \Psi + \hbar^2 \partial_{\phi,\phi}\Psi =0\, .
\end{equation}
The separability of the equation is obvious in these coordinates since the metric \eqref{ministach} has no dependence on $\phi$. By performing the standard separation of variables method with $\Psi(X,\phi)=\Xi(X)\Phi(\phi)$, we obtain the equations
\begin{align} \label{wdw1}
  &\hbar^2 \frac{d^2 \Phi}{d\phi^2} + \mathcal{E}^2 \Phi =0 \, , \\ \label{wdw2}
  &3 \hbar^2 \kappa \, V_0 \sqrt{1-2 X} \left(2 X (1-2 X) \frac{d^2\Xi}{dX^2}+(1-5 X) \frac{d \Xi}{dX}\right)+ \mathcal{E}^2 \Xi =0 \,  ,
\end{align}
where the $\mathcal{E}$ denotes the separation constant. The resulting solutions are
\begin{align} \label{Phioriginal}
  \Phi(\phi) & = C_1 \sin\left(\frac{\mathcal{E}}{\hbar} \phi \right) + C_2 \cos\left( \frac{\mathcal{E}}{\hbar} \phi\right) \\ \label{xioriginal}
  \Xi(X) & = C_3 \sin\left[ \frac{\mathcal{E}}{\hbar} \sqrt{\frac{2 X}{3 \kappa  V_0}}\, _2F_1\left(\frac{1}{2},\frac{3}{4};\frac{3}{2},2 X\right)\right] + C_4 \cos\left[\frac{\mathcal{E}}{\hbar}  \sqrt{\frac{2  X}{3 \kappa  V_0}} \, _2F_1\left(\frac{1}{2},\frac{3}{4},\frac{3}{2};2 X\right)\right] \, ,
\end{align}
where the $C_i$'s, $i=1,..,4$ represent the constants of integration. 

As noted in the classical solution, Eq. \eqref{solclphi} implies that the scalar field assumes values in a bounded region. We may thus impose the periodic boundary condition
\begin{equation}
  \Phi(0)=\Phi(\phi_{max}) = 0\, ,
\end{equation}
which results in the well-known orthonormal expressions
\begin{equation}\label{wavephi}
   \Phi_n(\phi) = \sqrt{\frac{2}{\phi_{\max}}} \sin\left( \frac{n \pi}{\phi_{\max}} \phi \right) , \quad n \in \mathbb{N}^* ,
\end{equation}
where $\phi_{max}$ is given by \eqref{maxphi} and the wave number $\mathcal{E}$ assumes the discrete values
\begin{equation}
  \mathcal{E}_n = \frac{n \pi \hbar }{\phi_{\max}} = \frac{n \sqrt{3\kappa \pi V_0} }{2}\frac{ \Gamma \left(3/4\right)}{\Gamma \left(5/4\right)} \hbar\, .
\end{equation}

The flatness of the metric \eqref{ministach} allows for the further simplification of the second equation involving $X$ by performing a transformation $X\rightarrow u$. If $X$ lies within region (R), i.e. $0 \leq X \leq 1/2$, the transformation reads
\begin{equation} \label{Xtou}
   \text{region (R)}: \quad u (X) = \frac{2 \sqrt{2} }{\sqrt{3\, \kappa } V_0^{3/2}} \sqrt{X} \, _2F_1\left(\frac{1}{2},\frac{3}{4},\frac{3}{2};2 X\right) \, .
\end{equation}
Then, \eqref{wdw2} becomes
\begin{equation} \label{wdw2b}
   4 \hbar^2 \frac{d^2\Xi^{(R)}}{du^2} +V_0^2 \mathcal{E}_n^2 \Xi^{(R)} =0 \, ,
\end{equation}
and its general solution is of course
\begin{equation} \label{solxiI}
  \Xi_n^{(R)}(u) = C_3 \cos\left( \frac{V_0 \mathcal{E}_n}{2 \hbar} u \right) + C_4 \sin\left( \frac{V_0 \mathcal{E}_n}{2 \hbar} u \right) \, .
\end{equation}
The variable $u$ from transformation \eqref{Xtou} is real as long as $X$ is in (R), that is between $0\leq X \leq 1/2$. The $u$ variable in this case varies in the range
\begin{equation}
   0\leq u \leq \sqrt{\frac{\pi }{3\, \kappa \, V_0^3}} \frac{ \Gamma \left(1/4\right)}{\Gamma \left(3/4\right)} \, .
\end{equation}
The introduction of the $u$ variable using \eqref{Xtou} is not accidental. It is chosen so that the mini-superspace metric $G_{ij}$ is brought to an obvious flat form with $|\det{G}_{ij}|=1$.

But, as we have seen in the classical analysis, there exists a Lorentzian metric as a solution for $X<0$, i.e. the region (L). The transformation \eqref{Xtou}, which we applied in region (R), is not appropriate now because we need to keep $u$ real. We thus introduce the transformation
\begin{align} \label{Xtou2}
   \text{region (L)}: \quad u (X) = \ima \frac{2 \sqrt{2 X} }{\sqrt{3\, \kappa } V_0^{3/2}}\, {}_2F_1\left(\frac{1}{2},\frac{3}{4},\frac{3}{2};2 X\right) \, .
\end{align}
With the previous variable definitions we have a continuous line in $u$ parted now in two connected regions
\begin{align}
  & \text{region (L)}: - u_{max}< u<0 \, , \\
  & \text{region (R)}: 0\leq u \leq u_{max} \, ,
\end{align}
where
\begin{equation}
  u_{max} =\sqrt{\frac{\pi }{3\, \kappa \, V_0^3}} \frac{ \Gamma \left(1/4\right)}{\Gamma \left(3/4\right)} .
\end{equation}
We leave the end point $-u_{max}$ in region (L) open since it corresponds to the limit $X\rightarrow -\infty$. 

The differential equation \eqref{wdw2} for region (L) now becomes
\begin{equation}
   4 \hbar^2 \frac{d^2\Xi^{(L)}_n}{du^2} - V_0^2 \mathcal{E}_n^2 \Xi^{(L)}_n =0 \, , 
\end{equation}
and its solution is
\begin{equation}
  \Xi_n^{(L)}(u) = C_5 \, e^{ \frac{V_0 \mathcal{E}_n}{2 \hbar} u } + C_6 \, e^{- \frac{V_0 \mathcal{E}_n}{2 \hbar} u } \, .
\end{equation}

By having obtained the two solutions for $X$, we need to take into account the relative junction conditions to guarantee the continuity of the wave function and its first derivative at the border. We thus enforce the following set of algebraic relations
\begin{equation} \label{janction}
  \Xi_n^{(L)}(0^{-}) = \Xi_n^{(R)}(0^{+}), \quad\quad \frac{d\Xi_n^{(L)}}{du}\Big|_{0^{-}} = \frac{d\Xi_n^{(R)}}{du}\Big|_{0^{+}} \,  ,
\end{equation}
which define two of the four involved constants of integration:
\begin{equation}
    C_3 = C_5+C_6, \quad C_4 = C_5 - C_6 \, .
\end{equation}
The normalization can fix one of the remaining two constants, either $C_5$ or $C_6$. For the final constant, an additional boundary condition is necessary. A valid option in this respect is to require that $\Psi\rightarrow 0$ as $X\rightarrow -\infty$, or, in terms of the $u$ variable, $u\rightarrow -u_{max}$. This results in the relation
\begin{equation} \label{boundary1}
  C_5=-\exp \left[\frac{\sqrt{\pi} \mathcal{E}_n \Gamma \left(1/4\right)}{\hbar \sqrt{3\kappa V_0 }  \Gamma \left(3/4\right)}\right] C_6 \, .
\end{equation}

\begin{figure}[ht]
  \centering
  \includegraphics[scale=0.7]{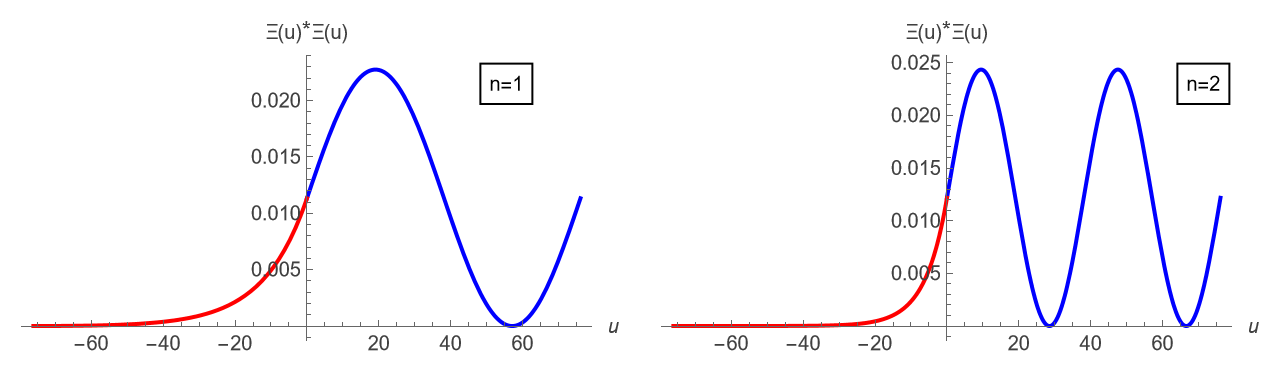}
  \caption{The probability density with respect to the $u$ variable for the states with $n=1$ and $n=2$. For the boundary condition \eqref{boundary1} we obtain a tunneling effect as the universe crosses the phantom line. The blue line corresponds to $0\geq w\geq-1$, the red to $w<-1$.} \label{fig4}
\end{figure}

\begin{figure}[ht]
  \centering
  \includegraphics[scale=0.7]{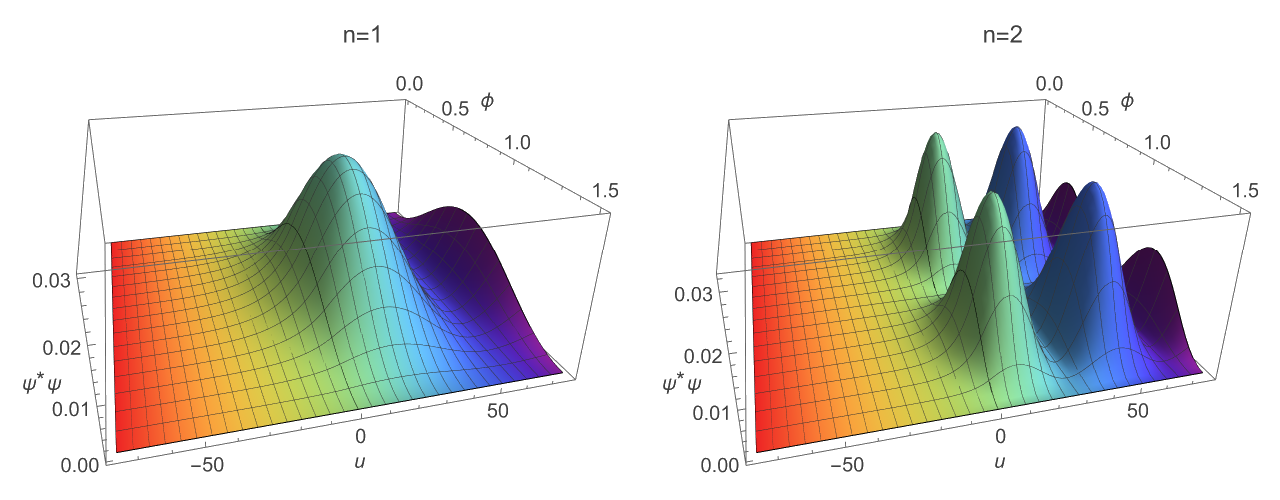}
  \caption{The total probability density including the $\phi$ variable for the states with $n=1$ and $n=2$. The crossing to the $w<-1$ region occurs at $u=0$.} \label{fig5}
\end{figure}

In Fig. \ref{fig4} we depict the total probability density $\Xi(u)^*\Xi(u)$ of the $u$ sector in both regions using boundary condition \eqref{boundary1}. The normalization is chosen so that $P_{(L)}+P_{(R)}=1$, where
\begin{equation}
  P_{(i)} = \int_{(i)} (\Xi_n^{(i)})^*\, \Xi_n^{(i)} du \, .
\end{equation}
The index $i$ here indicates one of each regions, $i=R$ for the right region and $i=L$ for the left. We observe a typical quantum tunneling towards the classically forbidden region $w<-1$, while for $0\geq w\geq-1$ we have an oscillatory behavior. In Fig. \ref{fig5} we present the corresponding graphs in three-dimensions, by adding the $\phi$ dependence of the full wave function.

As we previously mentioned, the boundary condition \eqref{boundary1} serves to send to zero the wave function when $X\rightarrow -\infty$; equivalently when $u\rightarrow -u_{max}$. This, however, results in a non-zero probability at the other extreme $X=1/2$, or $u=u_{max}$. Given that the classical energy density is given by Eq. \eqref{rhopeff} and it is of the form $\rho\propto (1-2X)^{-1/2}$, the corresponding integral representing the expectation value of the energy density,
\begin{equation} \label{Qenden}
  <\rho> = \int_\Omega \sqrt{-G} \rho(X) \Psi^* \Psi \, d\phi dX = \int_\Omega \rho(u) \Psi^* \Psi \, d\phi du \, ,
\end{equation}
where $\Omega$ denotes the full domain of variables $\phi$ and $X$, can be seen to diverge as the variable $u$ (or $X$) approaches the extreme limit of the right region (R), $X\rightarrow 1/2$. In this sense, the boundary condition $\underset{X\rightarrow-\infty}{\lim}\Psi=0$, leading to \eqref{boundary1} appears incompatible with the avoidance of the initial singularity. Remember that in the classical solution, the system starts from this point; the scale factor is zero at the instant corresponding to $X=1/2$.

\begin{figure}[ht]
  \centering
  \includegraphics[scale=0.7]{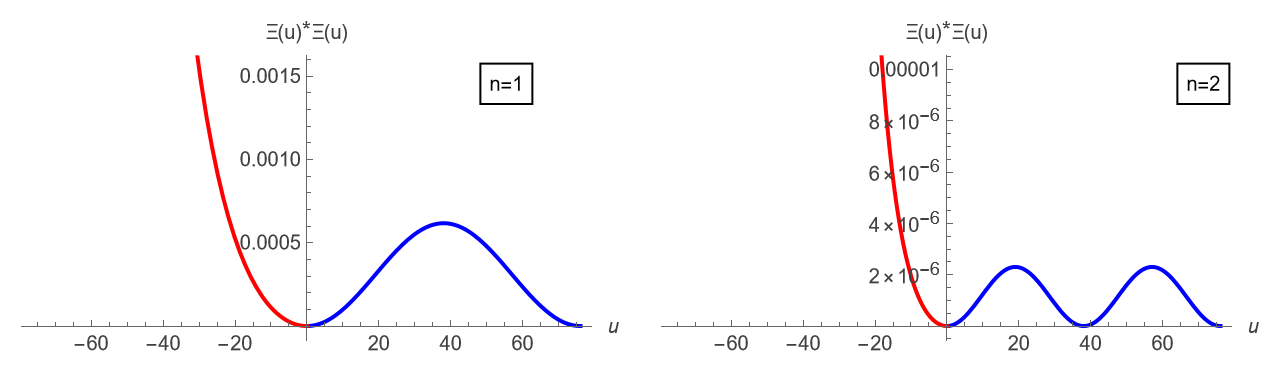}
  \caption{The probability density in the $u$ variable for the boundary condition $\Xi(u_{max})=0$. On the left (L) region the probability density stops at a finite value. The endpoint in $u$ there corresponds to $X\rightarrow-\infty$.} \label{fig6}
\end{figure}

To rectify this, let us impose $\Xi^{(R)}(u_{max})=0$, instead of using \eqref{boundary1}. This leads to the vanishing of the wave function at the classically problematic point (DeWitt boundary condition \cite{DeWitt1967}). Setting $u=u_{max}$ in \eqref{solxiI} we obtain
\begin{equation} \label{avoidsing}
  \Xi_n^{(R)}(u_{max}) = C_3 \cos(n \pi) + C_4 \sin(n \pi). 
\end{equation}
Due to the discretized values of $n$, this immediately implies $C_3=0$, which leaves only the sine branch of the solution. Imposing the same junction conditions as before, see Eq. \eqref{janction}, we now have $C_5=-C_6=C_4/2$. With only the sine remaining in $\Xi_n^{(R)}$, the singularity of the energy density can now be avoided. To see this, consider the integrand expressed in the $X$ variable
\begin{equation}
  \sqrt{-G(X)} \rho(X) (\Xi(X)^{(R)})^* \Xi(X)^{(R)} \propto \frac{1}{(1-2 X)^{5/4} \sqrt{X}} \left[\sin \left(\frac{\mathcal{E}_n \sqrt{2X}  \, _2F_1\left(\frac{1}{2},\frac{3}{4};\frac{3}{2};2 X\right)}{\hbar \sqrt{3 \kappa V_0}}\right)\right]^2 \, .
\end{equation}
The sine near the $X\rightarrow 1/2$ behaves as 
\begin{equation}
\begin{split}
 \sin \left(\frac{\mathcal{E}_n \sqrt{2X}  \, _2F_1\left(\frac{1}{2},\frac{3}{4};\frac{3}{2};2 X\right)}{\hbar \sqrt{3 \kappa V_0}}\right) & \sim \sin\left(n\pi- n \sqrt{\pi } \frac{\Gamma \left(3/4\right)}{\Gamma \left(5/4\right)}(1-2 X)^{1/4} \right) \\
 & \sim  n(-1)^{n+1} \sqrt{\pi } \frac{ \Gamma \left(3/4\right)}{\Gamma \left(5/4\right)} (1-2 X)^{1/4} \, .
 \end{split}
\end{equation}
This results in an integrand that, near the singularity, it becomes
\begin{equation}
  \sqrt{-G(X)}\rho(X) (\Xi(X)^{(R)})^* \Xi(X)^{(R)} \sim \sqrt{\frac{2}{3\kappa V_0}} \frac{ n^2\pi  \Gamma \left(3/4\right)^2}{\Gamma \left(5/4\right)^2\sqrt{X}(1-2 X)^{3/4} } \, .
\end{equation}
The order of the singularity is $p=3/4<1$, which renders the integrals involving $X=1/2$ convergent \cite{bookcalculus}. 
The resulting wave function in the $u$ variable, after normalization, can be observed in Fig. \ref{fig6}. We thus see that the avoidance of the classical singularity creates a node exactly at the point $u=0=X$, which corresponds to the phantom divide line $w=-1$ (see Eq. \eqref{weff}). Note that this does not imply a zero transition probability from one region to the other. This, due to the fact that there is a non-zero probability density on both regions next to this point and any probability integral containing it has also a non-zero value.

In the phantom region (L), the wave function becomes proportional to a hyperbolic sine function, and the corresponding probability density increases to a maximum finite value at the limit point $u=-u_{max}$. Notice that in the $X$ variable, this means that the probability density actually degreases as $X$ approaches $-\infty$. Due to the use of the Laplacian as our basic an operator, the probability density is invariant under coordinate transformations. Since the integral in $u$ has a finite value, the same holds for the integral in $X$. From the moment that, in this region, $X$ reaches $-\infty$ this can only happen if the probability density in the $X$ variable diminishes as $X\rightarrow -\infty$. It can be easily verified that in the $X$ variable, the probability density in the (L) region reads
\begin{equation}
  \sqrt{-G(X)} (\Xi^{(L)})^* \Xi^{(L)} \propto \frac{1}{(1-2 X)^{3/4} \sqrt{X}}\left[\sinh \left(\frac{\mathcal{E}_n \sqrt{-2X}  \, _2F_1\left(1/2,3/4;3/2;2 X\right)}{\hbar \sqrt{3 \kappa V_0}}\right)\right]^2 \, ,
\end{equation}
whose limit at $X \rightarrow -\infty$ is zero. 

The two distinct boundary value approaches imply different mean values for the $u$ variable
\begin{equation}
  < u > = \int_{-u_{max}}^0\!\! (\Xi^{(L)})^* \Xi^{(L)}  u \, du +\int_0^{u_{max}} \!\! (\Xi^{(R)})^* \Xi^{(R)} u  \, du .
\end{equation}
The mean values for the two boundary conditions $\Psi\rightarrow 0$ as $u\rightarrow -u_{max}$ and $\Psi\rightarrow 0$ as $u\rightarrow +u_{max}$  are calculated to be respectively 
\begin{equation}
   <u>^- =  \frac{2 n^2 \pi ^2 +\left(2 \pi ^2 n^2-1\right) \cosh (2 \, n\, \pi)-2 \, n \, \pi \sinh (2\, n\, \pi )+1}{n \, \sqrt{\kappa } V_0^{3/2} \Gamma \left(3/4\right)^2 \left(4 \pi  n \sinh ^2(n\, \pi)+\sinh (2\, n\, \pi )\right)} \sqrt{\frac{\pi }{6}}
\end{equation}
\begin{equation}
  <u>^+ = \frac{4\,  n^2 \pi ^2 -2 \, n\, \pi  \sinh (2 \,  n\, \pi )+\cosh (2\,  n\, \pi )-1}{n \, \sqrt{\kappa } \, V_0^{3/2} \Gamma \left(3/4\right)^2 \sinh (2 \,  n\,  \pi )} \sqrt{\frac{\pi }{6}} .
\end{equation}
The minus and plus superscripts denote the respective boundaries where the wave function is imposed to be zero. A few comments are in order here: The $<u>^-$ assumes positive values and $<u>^+$ negative. As $n$ increases, $<u>^-$ becomes more positive and $<u>^+$ more negative. By using the transformation relations \eqref{Xtou} and \eqref{Xtou2} to recover the values of $X$ corresponding to $<u>^-$ and $<u>^+$ we obtain: for $n=1$, $X^-\simeq 0.2$ and $X^+\simeq - 154 $ respectively. Of course, these are not the mean values in the $X$ variable as $u(X)$ is not a linear map. Nevertheless, due to the complicated expressions of the integrals in the $X$ variable, they can serve to set rough limits by invoking Jensen's inequalities for the mean values \cite{Jensen1,Jensen2}. 

In short, if a function is convex, i.e. $\frac{d^2 u}{dX^2}>0$, then $<X> \leq u^{-1}(<u(X)>)$ and if it is concave, $\frac{d^2 u}{dX^2}<0$, we have $<X> \geq u^{-1}(<u(X)>)$. The $X^+\simeq - 154 $ corresponds to a convex region of $u(X)$, which implies that the actual mean value $<X>^+$ is even more negative $<X>^+\leq -154$. As a result, the mean equation of state parameter, which is classically associated to the expansion rate through the deceleration parameter, would have a limit $<w>^+\leq -309$, which means that the boundary condition relating to the vanishing of the singularity in the $V=V_0$ case, pushes the system deep into the phantom region to unnaturally high values of $w$. On the other hand, the $X^-\simeq 0.2$ value lies to a point of inflexion. The latter suggest, that the actual mean value is very close to that number $<X>^-\simeq X^-\simeq 0.2$. This yields a more natural mean equation of state parameter  $<w>^-\simeq - 0.6$. Of course, we need to keep in mind that this is a toy model. What this correlation suggests, is that the boundary conditions, which may need to be taken differently in distinct potential scenarios since the classical singularity can occupy different points in the space of the introduced variables, may well govern the expected expansion rate in a given inflationary model.

\section{Conclusions}

In this work, we investigated the formulation of quantum cosmology for a general class of $k$-essence theories. We applied the Dirac–Bergmann algorithm to systematically identify all the constraints in the Hamiltonian formulation. The analysis led to a reduced Hamiltonian expressed in a purely quadratic form with no potential component.

As a specific example, we examined the case of the tachyonic field. In order to recover an exact quantum solution, we concentrated on a constant potential $V=V_0$. The latter serves as the relevant limit in the slow-roll approximation of inflationary scenarios that make use of a tachyonic field. By exploring the possible boundary conditions, we found that the avoidance of the classical singularity enforces a zero probability density exactly at the point corresponding to the $w=-1$ line. However, we expect that in the more realistic case of a non-constant potential function, $V(\phi)$, the situation could be substantially different. 

A striking result of our analysis is that the quantum setting is universal for any $k$-essence theory. There exist coordinates and choices of linear combinations of constraints that bring the corresponding Wheeler-DeWitt equation to the form of a two-dimensional d'Alembertian. We may, therefore, argue that the quantum description presented here will be preserved, as it is based on the same fundamental equations. Nevertheless, there exists a crucial difference, which is related to the choice of coordinates in which the metric of the configuration space takes its explicitly flat form. For the simple $V=V_0$ model, this was achieved through the introduction of the $u$ coordinate we described in our analysis.

To make it more illustrative, in our  $(u,\phi)$ coordinates the mini-supermetric $G_{ij}$ reads\footnote{Remember that the $u$ variable was chosen so that the measure function entering the probability integrals in the $u$ variable is$\sqrt{|G|}=\sqrt{|\det{G_{ij}}|}=1$.}
\begin{equation}
  G_{ij}(u,\phi) = \begin{pmatrix}
                     -\frac{V_0}{2} & 0 \\
                     0 & \frac{2}{V_0} 
                   \end{pmatrix} \, .
\end{equation}
Obviously, in a more complex problem where $V(\phi)$ is not constant, we would have to mix both variables and introduce a new set of coordinates $(X,\phi)\mapsto(u(X,\phi),v(X,\phi))$ in order to achieve the above form for the metric. This severely affects the boundary conditions, which in the $V=V_0$ case imposed a node at the phantom divide in order to avoid the singularity. First of all, it may happen that no variable is bounded, as was the case with $\phi$ for the $V=V_0$ model. This would lead to a continuous spectrum for $\mathcal{E}$. The discreteness of the latter is one of the reasons why we obtained a node at $u=0$ when $\Psi$ is also zero at the $u=u_{max}$ (equivalently $X=1/2$). Secondly, within the set $(u,v)$ it is not straightforward to determine how the boundaries may be connected with the actual classical singularity in the model. This is something which needs to be checked in a case-by-case scenario, depending on the domains of the variables given by the classical solutions and the transformations leading to the $u,v$ variables. 

The relation of the $u$ and $v$ coordinates to the original $\phi$ and $X$ variables, as well as the form of the classical solution, can severely affect the positioning of the classical singularity. In the toy model under consideration it corresponded to $X=0$ (in actual inflationary scenarios these would not be the starting point), and through the chosen transformation it was mapped to $u\rightarrow +u_{max}$. As we have seen, extremely different results are obtained depending on where the wave function is enforced to vanish, which suggests that not all quantum boundary conditions yield a physically plausible classical universe.

Although the situation can be quite involved even for simple potentials, in a future investigation, we plan to extend this analysis to reveal a  possible deeper interaction between singularity avoidance and the qualitative behavior of the probability density, especially in what regards phantom crossing. An additional interesting direction would be the investigation of singular points of the minisuperspace, possibly related to phase-space degeneracies observed in theories with nontypical kinetic terms \cite{fin1,fin2,fin3,fin4}.

\begin{acknowledgments}
AP acknowledges the support from FONDECYT Grant 1240514.
\end{acknowledgments}


\begin{thebibliography}{120}


\bibitem{Fischetti:1979ue}
M.V.~Fischetti, J.B.~Hartle and B.L.~Hu,
``Quantum Effects in the Early Universe. I. Influence of Trace Anomalies on Homogeneous, Isotropic, Classical Geometries,''
\href{https://doi.org/10.1103/PhysRevD.20.1757}{Phys. Rev. D \textbf{20} (1979), 1757-1771}

\bibitem{Hartle:1979uf}
J.B.~Hartle and B.L.~Hu,
``Quantum Effects in the Early Universe. II. Effective Action for Scalar Fields in Homogeneous Cosmologies with Small Anisotropy,''
\href{https://doi.org/10.1103/PhysRevD.20.1772}{Phys. Rev. D \textbf{20} (1979), 1772-1782}

\bibitem{Hartle:1980nn}
J.B.~Hartle and B.L.~Hu,
``Quantum Effects in the Early Universe. III. Dissipation of Anisotropy by Scalar Particle Production,''
\href{https://doi.org/10.1103/PhysRevD.21.2756}{Phys. Rev. D \textbf{21} (1980), 2756-2769}

\bibitem{Hartle:1980kz}
J.B.~Hartle,
``Quantum Effects in the Early Universe. IV. Nonlocal Effects in Particle Production in Anisotropic Models,''
\href{https://doi.org/10.1103/PhysRevD.22.2091}{Phys. Rev. D \textbf{22} (1980), 2091-2095}

\bibitem{Kiefer:2025udf}
C.~Kiefer,
``Quantum Gravity,''
Oxford University Press (2025)

\bibitem{DeWitt1967} B.S. DeWitt, ``Quantum theory of gravity. I. The canonical theory,'' \href{https://doi.org/10.1103/PhysRev.160.1113}{Phys. Rev. \textbf{160}, 1113 (1967)}
    
\bibitem{Kuchar} K. Kucha\v{r}, ``Canonical Quantization of Gravity,'' in ``Relativity, Astrophysics and Cosmology,'' Edts. W. Israel, D. Reidel Publishing Company, Dordrecht, Boston, (1973)


\bibitem{Hawking:1978jz}
S.W.~Hawking,
``Quantum Gravity and Path Integrals,''
\href{https://doi.org/10.1103/PhysRevD.18.1747}{Phys. Rev. D \textbf{18} (1978), 1747-1753}

\bibitem{Teitelboim:1981ua}
C.~Teitelboim,
``Quantum Mechanics of the Gravitational Field,''
\href{https://doi.org/10.1103/PhysRevD.25.3159}{Phys. Rev. D \textbf{25} (1982), 3159}


\bibitem{Ashtekar:2008zu}
A.~Ashtekar,
``Loop Quantum Cosmology: An Overview,''
\href{https://doi.org/10.1007/s10714-009-0763-4}{Gen. Rel. Grav. \textbf{41} (2009), 707-741}

\bibitem{Rovelli} C. Rovelli, ``Loop quantum gravity: the first twenty five years,'' \href{https://doi.org/10.1088/0264-9381/28/15/153002}{Class. Quantum Grav. \textbf{28} (2011), 153002}


\bibitem{Perez:2012wv}
A.~Perez,
``The Spin Foam Approach to Quantum Gravity,''
\href{https://doi.org/10.12942/lrr-2013-3}{Living Rev. Rel. \textbf{16} (2013), 3}

\bibitem{Asante} S.K. Asante, B. Dittrich and H.M. Haggard, ``Effective Spin Foam Models for Four-Dimensional Quantum Gravity,'' \href{https://doi.org/10.1103/PhysRevLett.125.231301}{Phys. Rev. Lett. \textbf{125} (2020), 231301}


\bibitem{Vilenkin:1994rn}
A.~Vilenkin,
``Approaches to quantum cosmology,''
\href{https://doi.org/10.1103/PhysRevD.50.2581}{Phys. Rev. D \textbf{50} (1994), 2581-2594}


\bibitem{Bojowald:2015iga}
M.~Bojowald,
``Quantum cosmology: a review,''
\href{https://doi.org/10.1088/0034-4885/78/2/023901}{Rept. Prog. Phys. \textbf{78} (2015), 023901}


\bibitem{Kiefer:2008sw}
C.~Kiefer and B.~Sandhoefer,
``Quantum cosmology,''
\href{https://doi.org/10.1515/zna-2021-0384}{Z. Naturforsch. A \textbf{77} (2022) no.6, 543-559}

\bibitem{Reuter} M. Reuter and F. Saueressig, ``Asymptotic Safety, Fractals, and Cosmology,'' in ``Quantum Gravity and Quantum Cosmology,'' Edts. G. Calcagni, L. Papantonopoulos, G. Siopsis, N. Tsamis, Lecture Notes in Physics, vol 863. Springer, Berlin, Heidelberg (2013)
    
\bibitem{Surya} S. Surya, ``The causal set approach to quantum gravity,'' \href{https://doi.org/10.1007/s41114-019-0023-1}{Living Rev. Relativ. \textbf{22} (2019), 5}
    
\bibitem{Calcagni} G. Calcagni, ``Fractal Universe and Quantum Gravity,'' \href{https://doi.org/10.1103/PhysRevLett.104.251301}{Phys. Rev. Lett. \textbf{104} (2010), 251301}

\bibitem{Hartle:1983ai}
J.B.~Hartle and S.W.~Hawking,
``Wave Function of the Universe,''
\href{https://doi.org/10.1103/PhysRevD.28.2960}{Phys. Rev. D \textbf{28} (1983), 2960-2975}


\bibitem{Kiefer:2010zzb}
C.~Kiefer, ``Can singularities be avoided in quantum cosmology?,''
\href{https://doi.org/10.1002/andp.201052203-510}{Annalen Phys. \textbf{522} (2010), 211-218}

\bibitem{Kiefer:2022} C. Kiefer and B. Sandhoefer, ``Quantum Cosmology,'' \href{https://doi.org/10.1515/zna-2021-0384}{Z. Naturforsch. A \textbf{77} (2022), 543-559}

\bibitem{Vakili} B. Vakili, ``Classical and quantum dynamics of a perfect fluid scalar-metric cosmology,'' \href{https://doi.org/10.1016/j.physletb.2010.04.007}{Phys.Lett.B \textbf{688} (2010), 129-136}

\bibitem{Zampeli:2015ojr}
A.~Zampeli, T.~Pailas, P.~A.~Terzis and T.~Christodoulakis,
``Conditional symmetries in axisymmetric quantum cosmologies with scalar fields and the fate of the classical singularities,''
\href{https://doi.org/10.1088/1475-7516/2016/05/066}{JCAP \textbf{05} (2016), 066}

\bibitem{Misner} C.W. Minser, ``Quantum Cosmology. I,'' \href{https://doi.org/10.1103/PhysRev.186.1319}{Phys. Rev. \textbf{186} (1969), 1319}

\bibitem{Berger} B.K. Berger, ``Quantum cosmology: Exact solution for the Gowdy T3 model,'' \href{https://doi.org/10.1103/PhysRevD.11.2770}{Phys. Rev. D \textbf{11} (1975), 2770}

\bibitem{Lemos} N.A. Lemos, ``Singularities in a Scalar Field Quantum Cosmology,'' \href{https://doi.org/10.1103/PhysRevD.53.4275}{Phys. Rev. D \textbf{53} (1996), 4275}

\bibitem{Vilenkin2002} A. Vilenkin, ``Quantum cosmology and eternal inflation,'' (2002) [\href{https://arxiv.org/abs/gr-qc/0204061}{arXiv:gr-qc/0204061}]


\bibitem{Vilenkin1988} A. Vilenkin, ``Quantum cosmology and the initial state of the Universe,'' \href{https://doi.org/10.1103/PhysRevD.37.888}{Phys. Rev. D \textbf{37}, 888 (1988)}

\bibitem{Halliwell1990} J.J. Halliwell, ``Introductory lectures on quantum cosmology,'' (1990) [\href{https://arxiv.org/abs/0909.2566}{arXiv:0909.2566 [gr-qc]}]

\bibitem{Halliwell1988} J.J. Halliwell, ``Derivation of the Wheeler–DeWitt equation from a path integral for minisuperspace models,'' \href{https://doi.org/10.1103/PhysRevD.38.2468}{Phys. Rev. D \textbf{38}, 2468 (1988)}

\bibitem{Linde1990} A.D. Linde, ``Inflation and Quantum Cosmology,'' Academic Press, Boston, San Diego, New York, (1990)

\bibitem{Handley2021} W. Handley and T. Gill, ``Constraining quantum initial conditions before inflation,'' \href{https://doi.org/10.1103/PhysRevD.104.063532}{Phys. Rev. D \textbf{104}, 063532 (2021)} 
    
\bibitem{QC1} S. Gielen and L. Menéndez-Pidal, ``Singularity resolution depends on the clock,'' \href{https://doi.org/10.1088/1361-6382/abb14f}{Class. Quantum Grav. \textbf{37} (2020) 205018}
    
\bibitem{QC2} E. Alesci, G. Botta, F. Cianfrani and S. Liberati, ``Cosmological singularity resolution from quantum gravity: The emergentbouncing universe,'' \href{https://doi.org/10.1103/PhysRevD.96.046008}{Phys. Rev. D \textbf{96} (2017), 046008}
    
\bibitem{QC3} H. Bergeron, J. de Cabo Martin, J.-P. Gazeau and P. Malkiewicz, ``Can a quantum mixmaster universe undergo a spontaneous inflationary phase?,'' \href{https://doi.org/10.1103/PhysRevD.108.043534}{Phys. Rev. D \textbf{108} (2023), 043534}
    
\bibitem{QC4} A. Burton-Villalobos, G. Otalora, M. Gonzalez-Espinoza and Y. Leyva, ``Classical and quantum cosmology of $f(R)$ gravity's rainbow in Schutz's formalism,'' \href{https://doi.org/10.1140/epjc/s10052-025-14454-w}{Eur. Phys. J. C \textbf{85} (2025), 716}
    
\bibitem{QC5} O. Janssen, ``Slow-roll approximation in quantum cosmology,'' \href{https://doi.org/10.1088/1361-6382/abe143}{Class. Quantum Grav. \textbf{38} (2021), 095003}
    
\bibitem{Matone} M. Matone and N. Dimakis, ``Quantum mechanics from general relativity and the quantum Friedmann equation,'' \href{https://doi.org/10.1140/epjc/s10052-025-14930-3}{Eur. Phys. J. C \textbf{85} (2025), 1206}
    
\bibitem{Dimakistime} N. Dimakis, ``Time operator from parametrization invariance and implications for cosmology,'' \href{https://doi.org/10.1103/PhysRevD.111.083536}{Phys.Rev.D \textbf{111} (2025), 083536}
    
\bibitem{QCextra1} G. Kouniatalis, ``The Wave Function of the Universe and Inflation,'' \href{https://arxiv.org/abs/2510.04775}{arXiv:2510.04775 [gr-qc]}


\bibitem{Bohm:1951xw}
D.~Bohm,
``A Suggested interpretation of the quantum theory in terms of hidden variables. I.,
\href{https://doi.org/10.1103/PhysRev.85.166}{Phys. Rev. \textbf{85} (1952), 166-179}

\bibitem{Bohm:1951xx}
D.~Bohm,
``A Suggested interpretation of the quantum theory in terms of hidden variables. II.,''
\href{https://doi.org/10.1103/PhysRev.85.180}{Phys. Rev. \textbf{85} (1952), 180-193}


\bibitem{Pinto-Neto:2018zvn}
N.~Pinto-Neto and W.~Struyve,
``Bohmian quantum gravity and cosmology,'' in “Applied Bohmian Mechanics,” Eds. X. O. Pladevall, J. Mompart, Jenny Stanford Publishing, New York (2019), DOI: \href{https://doi.org/10.1201/9780429294747}{10.1201/9780429294747} 
[\href{https://doi.org/10.48550/arXiv.1801.03353}{arXiv:1801.03353 [gr-qc]}]


\bibitem{Vicente:2021abv}
G.S.~Vicente,
``Quantum Hǒrava-Lifshitz cosmology in the de Broglie{\textendash}Bohm interpretation,''
\href{https://doi.org/10.1103/PhysRevD.104.103525}{Phys. Rev. D \textbf{104} (2021), 103525}

\bibitem{Paliathanasis:2017ocj}
A.~Paliathanasis,
``Dust fluid component from Lie symmetries in Scalar field Cosmology,''
\href{https://doi.org/10.1142/S0217732317502066}{Mod. Phys. Lett. A \textbf{32} (2017), 1750206}

\bibitem{Paliathanasis:2018ixu}
A.~Paliathanasis, A.~Zampeli, T.~Christodoulakis and M.T.~Mustafa,
``Quantization of the Szekeres System,''
\href{https://doi.org/10.1088/1361-6382/aac227}{Class. Quant. Grav. \textbf{35} (2018), 125005}



\bibitem{Pinto-Neto:2021gcl}
N.~Pinto-Neto,
``Bouncing Quantum Cosmology,''
\href{https://doi.org/10.3390/universe7040110}{Universe \textbf{7} (2021), 110}

\bibitem{Vicente:2023hba}
G.S.~Vicente, R.O.~Ramos and V.N.~Magalh{\~a}es,
``Bouncing and inflationary dynamics in quantum cosmology in the de Broglie{\textendash}Bohm interpretation,''
\href{https://doi.org/10.1103/PhysRevD.108.023517}{Phys. Rev. D \textbf{108} (2023), 023517}

\bibitem{Basilakos:2025hyk}
S.~Basilakos, G.~Kouniatalis, E.N.~Saridakis and C.~Tzerefos,
``Bohmian Quantum Cosmology from the Wheeler-DeWitt Equation,''
[\href{https://doi.org/10.48550/arXiv.2512.18818}{arXiv:2512.18818 [gr-qc]}]

\bibitem{Starobinsky1980} A.A. Starobinsky, ``A new type of isotropic cosmological models without singularity,'' \href{https://doi.org/10.1016/0370-2693(80)90670-X}{Phys. Lett. B \textbf{91}, 99 (1980)}

\bibitem{Guth1981} A.H. Guth, ``Inflationary universe: A possible solution to the horizon and flatness problems,'' \href{https://doi.org/10.1103/PhysRevD.23.347}{Phys. Rev. D \textbf{23}, 347 (1981)}

\bibitem{Linde1982} A.D. Linde, ``A new inflationary universe scenario: A possible solution of the horizon, flatness, homogeneity, isotropy and primordial monopole problems,'' \href{https://doi.org/10.1016/0370-2693(82)91219-9}{Phys. Lett. B \textbf{108}, 389 (1982)}. 

\bibitem{AlbrechtSteinhardt1982} A. Albrecht and P.J. Steinhardt, ``Cosmology for grand unified theories with radiatively induced symmetry breaking,'' \href{https://doi.org/10.1103/PhysRevLett.48.1220}{Phys. Rev. Lett. \textbf{48}, 1220 (1982)}.

\bibitem{AlbrechtReheating1982} A. Albrecht, P.J. Steinhardt, M. S. Turner, and F. Wilczek, ``Reheating an inflationary universe,'' \href{https://doi.org/10.1103/PhysRevLett.48.1437}{Phys. Rev. Lett. \textbf{48}, 1437 (1982)}

\bibitem{Kofman1994} L. Kofman, A.D. Linde, and A.A. Starobinsky, “Reheating after inflation,” \href{https://doi.org/10.1103/PhysRevLett.73.3195}{Phys. Rev. Lett. \textbf{73}, 3195 (1994)} [\href{https://arxiv.org/abs/hep-th/9405187}{arXiv:hep-th/9405187}].  

\bibitem{Kofman1997} L. Kofman, A.D. Linde, and A.A. Starobinsky, ``Towards the theory of reheating after inflation,'' \href{https://doi.org/10.1103/PhysRevD.56.3258}{Phys. Rev. D \textbf{56}, 3258 (1997)}


\bibitem{Barrow:1995xb}
J.D.~Barrow and P.~Parsons,
``Inflationary models with logarithmic potentials,''
\href{https://doi.org/10.1103/PhysRevD.52.5576}{Phys. Rev. D \textbf{52} (1995), 5576-5587}

\bibitem{Lidsey:1991zp}
J.E.~Lidsey,
``The Scalar field as dynamical variable in inflation,''
\href{https://doi.org/10.1016/0370-2693(91)90550-A}{Phys. Lett. B \textbf{273} (1991), 42-46}

\bibitem{Ford:1989me}
L.H.~Ford,
``Inflation Driven by a Vector Field,''
\href{https://doi.org/10.1103/PhysRevD.40.967}{Phys. Rev. D \textbf{40} (1989), 967}

\bibitem{Heusler:1991ep}
M.~Heusler,
``Anisotropic asymptotic behavior in chaotic inflation,''
\href{https://doi.org/10.1016/0370-2693(91)91359-4}{Phys. Lett. B \textbf{253} (1991), 33-37}

\bibitem{Kim:2010fq}
Y.~Kim and S.C.~Park,
``Hyperbolic Inflation,''
\href{https://doi.org/10.1103/PhysRevD.83.066009}{Phys. Rev. D \textbf{83} (2011), 066009}

\bibitem{Lyth:1998xn}
D.H.~Lyth and A.~Riotto,
``Particle physics models of inflation and the cosmological density perturbation,''
\href{https://doi.org/10.1016/S0370-1573(98)00128-8}{Phys. Rept. \textbf{314} (1999), 1-146}

\bibitem{Adams:1997de}
J.A.~Adams, G.G.~Ross and S.~Sarkar,
``Multiple inflation,''
\href{https://doi.org/10.1016/S0550-3213(97)00431-8}{Nucl. Phys. B \textbf{503} (1997), 405-425}

\bibitem{Thomas:1995dq}
S.D.~Thomas,
``Moduli inflation from dynamical supersymmetry breaking,''
\href{https://doi.org/10.1016/0370-2693(95)00417-J}{Phys. Lett. B \textbf{351} (1995), 424-430}


\bibitem{Binetruy:1986ss}
P.~Bin{\'e}truy and M.K.~Gaillard,
``Candidates for the Inflaton Field in Superstring Models,''
\href{https://doi.org/10.1103/PhysRevD.34.3069}{Phys. Rev. D \textbf{34} (1986), 3069-3083}

\bibitem{Herrera:2024rcm}
R.~Herrera, M.~Shokri and J.~Sadeghi,
``Constant-roll inflation with a complex scalar field,''
\href{https://doi.org/10.1016/j.aop.2024.169705}{Annals Phys. \textbf{467} (2024), 169705}

\bibitem{Luongo:2024opv}
O.~Luongo and T.~Mengoni,
``Generalized K-essence inflation in Jordan and Einstein frames,''
\href{https://doi.org/10.1088/1361-6382/ad3ac9}{Class. Quant. Grav. \textbf{41} (2024), 105006}

\bibitem{Luongo:2023aaq}
O.~Luongo and T.~Mengoni,
``Quasi-quintessence inflation with non-minimal coupling to curvature in the Jordan and Einstein frames,''
[\href{https://doi.org/10.48550/arXiv.2309.03065}{arXiv:2309.03065 [gr-qc]}]

\bibitem{DAgostino:2022fcx}
R.~D'Agostino, O.~Luongo and M.~Muccino,
``Healing the cosmological constant problem during inflation through a unified quasi-quintessence matter field,''
\href{https://doi.org/10.1088/1361-6382/ac8af2}{Class. Quant. Grav. \textbf{39} (2022), 195014}

\bibitem{Barrow:2016qkh}
J.D.~Barrow and A.~Paliathanasis,
``Observational Constraints on New Exact Inflationary Scalar-field Solutions,''
\href{https://doi.org/10.1103/PhysRevD.94.083518}{Phys. Rev. D \textbf{94} (2016), 083518}


\bibitem{Linde:1983mx}
A.D.~Linde,
``Quantum Creation of the Inflationary Universe,''
\href{https://doi.org/10.1007/BF02790571}{Lett. Nuovo Cim. \textbf{39} (1984), 401-405}


\bibitem{Linde:1995ck}
A.D.~Linde,
``Quantum cosmology and the structure of inflationary universe,''
[\href{https://doi.org/10.48550/arXiv.gr-qc/9508019}{arXiv:gr-qc/9508019 [gr-qc]}]

\bibitem{Khoury:2022ish}
J.~Khoury and S.S.C.~Wong,
``Bayesian reasoning in eternal inflation: A solution to the measure problem,''
\href{https://doi.org/10.1103/PhysRevD.108.023506}{Phys. Rev. D \textbf{108} (2023), 023506}

\bibitem{Jalalzadeh:2022dlj}
S.~Jalalzadeh, A.~Mohammadi and D.~Demir,
``A quantum cosmology approach to cosmic coincidence and inflation,''
\href{https://doi.org/10.1016/j.dark.2023.101227}{Phys. Dark Univ. \textbf{40} (2023), 101227}

\bibitem{Martin:2013tda}
J.~Martin, C.~Ringeval and V.~Vennin,
``Encyclop{\ae}dia Inflationaris: Opiparous Edition,''
\href{https://doi.org/10.1016/j.dark.2024.101653}{Phys. Dark Univ. \textbf{46} (2024), 101653}


\bibitem{Balkenhol2025} L. Balkenhol \textit{et al.}, ``Inflation at the End of 2025: Constraints on $r$ and $n_s$ Using the Latest CMB and BAO Data,'' \textit{Open J. Astrophys.} (2025) [\href{https://arxiv.org/abs/2512.10613}{arXiv:2512.10613 [astro-ph.CO]}]

\bibitem{ArmendarizPicon1999} C. Armendáriz-Picón, T. Damour and V.F. Mukhanov, ``$k$-Inflation,'' \href{https://doi.org/10.1016/S0370-2693(99)00603-6}{Phys. Lett. B \textbf{458}, 209 (1999)} [\href{https://arxiv.org/abs/hep-th/9904075}{arXiv:hep-th/9904075}]

\bibitem{ASen1999}
A.~Sen,
``Non-BPS states and Branes in string theory,'' \href{https://doi.org/10.1088/0264-9381/17/5/334}{Class. Quantum Grav. \textbf{17}, 1251 (2000)}
[\href{https://arxiv.org/abs/hep-th/9904207}{arXiv:hep-th/9904207}]


\bibitem{DESI2025_DR2} DESI Collaboration, ``DESI DR2 results. II. Measurements of baryon acoustic oscillations and cosmological constraints,'' \href{https://doi.org/10.1103/tr6y-kpc6}{Phys. Rev. D \textbf{112}, 083515 (2025)} 


\bibitem{Li:2025cxn}
C.~Li, J.~Wang, D.~Zhang, E.N.~Saridakis and Y.F.~Cai,
``Quantum gravity meets DESI: dynamical dark energy in light of the trans-Planckian censorship conjecture,''
\href{https://doi.org/10.1088/1475-7516/2025/08/041}{JCAP \textbf{08} (2025), 041}

\bibitem{Oriti:2021rvm}
D.~Oriti and X.~Pang,
``Phantom-like dark energy from quantum gravity,''
\href{https://doi.org/10.1088/1475-7516/2021/12/040}{JCAP \textbf{12} (2021), 040}

\bibitem{Paliathanasis:2025dcr}
A.~Paliathanasis,
``Dark energy within the generalized uncertainty principle in light of DESI DR2,''
\href{https://doi.org/10.1088/1475-7516/2025/09/067}{JCAP \textbf{09} (2025), 067}

\bibitem{Paliathanasis:2025kmg}
A.~Paliathanasis, G.~Leon, Y.~Leyva, G.G.~Luciano and A.~Abebe,
``Challenging {\ensuremath{\Lambda}}CDM with higher-order GUP corrections,''
\href{https://doi.org/10.1016/j.jheap.2025.100533}{JHEAp \textbf{51} (2026), 100533}

\bibitem{Singh:2019hhi}
T.P.~Singh,
``Dark energy as a large scale quantum gravitational phenomenon,''
\href{https://doi.org/10.1142/S0217732320501953}{Mod. Phys. Lett. A \textbf{35} (2020), 2050195}


\bibitem{Dimakis:2020tzc}
N.~Dimakis and A.~Paliathanasis,
``Crossing the phantom divide line as an effect of quantum transitions,''
\href{https://doi.org/10.1088/1361-6382/abdaf6}{Class. Quant. Grav. \textbf{38} (2021), 075016}

\bibitem{ArmendarizPicon2001} C. Armendáriz-Picón, V. F. Mukhanov, and P. J. Steinhardt, ``Essentials of $k$-essence,'' \href{https://doi.org/10.1103/PhysRevD.63.103510}{Phys. Rev. D \textbf{63}, 103510 (2001)} [\href{https://arxiv.org/abs/astro-ph/0006373}{arXiv:astro-ph/0006373}].

\bibitem{Bahamonde2015}
S.~Bahamonde, C.~G.~B{\"o}hmer, F.~S.~N.~Lobo and D.~S{\'a}ez-G{\'o}mez,
``Generalized $f(R,\phi,X)$ Gravity and the Late-Time Cosmic Acceleration,''
\href{https://doi.org/10.3390/universe1020186}{Universe \textbf{1} (2015) no.2, 186-198} [\href{https://arxiv.org/abs/1506.07728}
{arXiv:1506.07728 [gr-qc]}].
\bibitem{Bose:2008ew}
N.~Bose and A.S.~Majumdar,
``A $k$-essence Model Of Inflation, Dark Matter and Dark Energy,''
\href{https://doi.org/10.1103/PhysRevD.79.103517}{Phys. Rev. D \textbf{79} (2009), 103517}

\bibitem{Bose:2009kc}
N.~Bose and A.S.~Majumdar,
``Unified Model of $k$-Inflation, Dark Matter {\&} Dark Energy,''
\href{https://doi.org/10.1103/PhysRevD.80.103508}{Phys. Rev. D \textbf{80} (2009), 103508}

\bibitem{Saa} S. Chakraborty, S.E. Jorás and A. Saa, ``Unified dynamical system formulations for $f(R,\phi,X)$ gravity with applications to nonminimal derivative coupling and $R^2$-Higgs inflation,'' [\href{https://doi.org/10.48550/arXiv.2512.16176}{arXiv:2512.16176 [gr-qc]}]


\bibitem{Csillag:2025gnz}
L.~Csillag and E.~Jensko,
``Geometric formulation of $k$-essence and late-time acceleration,''
[\href{https://doi.org/10.48550/arXiv.2505.15975}{arXiv:2505.15975 [gr-qc]}]


\bibitem{Bagla:2002yn}
J.S.~Bagla, H.K.~Jassal and T.~Padmanabhan,
``Cosmology with tachyon field as dark energy,''
\href{https://doi.org/10.1103/PhysRevD.67.063504}{Phys. Rev. D \textbf{67} (2003), 063504}

\bibitem{Xiong:2007ak}
H.H.~Xiong and J.Y.~Zhu,
``Tachyon field in loop quantum cosmology: Inflation and evolution picture,''
\href{https://doi.org/10.1103/PhysRevD.75.084023}{Phys. Rev. D \textbf{75} (2007), 084023}

\bibitem{Shi:2021dpi}
J.~Shi and J.P.~Wu,
``Dynamics of $k$-essence in loop quantum cosmology,''
\href{https://doi.org/10.1088/1674-1137/abe111}{Chin. Phys. C \textbf{45} (2021), 045104}


\bibitem{Ferraro:2018tpu}
R.~Ferraro and M.J.~Guzm{\'a}n,
``Hamiltonian formalism for $f(T)$ gravity,''
\href{https://doi.org/10.1103/PhysRevD.97.104028}{Phys. Rev. D \textbf{97} (2018), 104028}

\bibitem{Tomonari:2023wcs}
K.~Tomonari and S.~Bahamonde,
``Dirac{\textendash}Bergmann analysis and degrees of freedom of coincident $f(Q)$-gravity,''
\href{https://doi.org/10.1140/epjc/s10052-024-12677-x}{Eur. Phys. J. C \textbf{84} (2024), 349} 
[erratum: \href{https://doi.org/10.1140/epjc/s10052-024-12813-7}{Eur. Phys. J. C \textbf{84} (2024), 508}]

\bibitem{Dimakis:2021gby}
N.~Dimakis, A.~Paliathanasis and T.~Christodoulakis,
``Quantum cosmology in $f(Q)$ theory,''
\href{https://doi.org/10.1088/1361-6382/ac2b09}{Class. Quant. Grav. \textbf{38} (2021), 225003}

\bibitem{Dimakis:2023oje}
N.~Dimakis, A.~Paliathanasis and T.~Christodoulakis,
``Exploring quantum cosmology within the framework of teleparallel $f(T)$ gravity,''
\href{https://doi.org/10.1103/PhysRevD.109.024031}{Phys. Rev. D \textbf{109} (2024) no.2, 024031}

\bibitem{DAmbrosio:2023asf}
F.~D'Ambrosio, L.~Heisenberg and S.~Zentarra,
``Hamiltonian Analysis of $f(\mathbb {Q})$ Gravity and the Failure of the Dirac{\textendash}Bergmann Algorithm for Teleparallel Theories of Gravity,''
\href{https://doi.org/10.1002/prop.202300185}{Fortsch. Phys. \textbf{71} (2023), 2300185}

\bibitem{Tomonari:2024ybs}
K.~Tomonari and D.~Blixt,
``Degrees of freedom of new general relativity: Type 2, type 3, type 5, and type 8,''
\href{https://doi.org/10.1103/4ggt-6nd4}{Phys. Rev. D \textbf{112} (2025), 084052}


\bibitem{Dirac} P.A.M. Dirac, ``Generalized Hamiltonian Dynamics,'' \href{https://doi.org/10.4153/CJM-1950-012-1}{Canad. J. Math \textbf{2}, (1950) 129}.

\bibitem{Bergmann} J. Anderson and P. Bergmann, ``Constraints in Covariant Field Theories,''  \href{https://doi.org/10.1103/PhysRev.83.1018}{Phys. Rev. \textbf{83}, (1951) 1018}.



\bibitem{Diracbook} P.A.M. Dirac, ``Lectures on Quantum Mechanics,'' Yeshiva University Press, New York (1964).

\bibitem{Sund} K. Sundermeyer, ``Constrained Dynamics,'' Springer-Verlag, Berlin, Heidelberg (1982).

\bibitem{Sen1} A. Sen, ``Rolling Tachyon,'' \href{https://doi.org/10.1088/1126-6708/2002/04/048}{JHEP 0204, (2002) 048} 
    
\bibitem{Sen2} A. Sen, ``Tachyon Matter,'' \href{https://doi.org/10.1088/1126-6708/2002/07/065}{JHEP 0207, (2002) 065} 
    
\bibitem{Gibbons} G.W. Gibbons, ``Cosmological Evolution of the Rolling Tachyon,'' \href{https://doi.org/10.1016/S0370-2693\%2802\%2901881-6}{Phys. Lett. B \textbf{537}, (2002) 1} 
    
\bibitem{inflt1} M. Fairbairn and M.H.G. Tytgat, ``Inflation from a Tachyon Fluid?'' \href{https://doi.org/10.1016/S0370-2693\%2802\%2902638-2}{Phys. Lett. B \textbf{546}, (2002) 1} 


\bibitem{Gibbons2} G.W. Gibbons, ``Thoughts on Tachyon Cosmology,'' \href{https://doi.org/10.1088/0264-9381/20/12/301}{Class. Quantum Grav. \textbf{20}, (2003) S321-S346} 
    
\bibitem{bookcalculus} G.B. Folland, ``Advanced Calculus,'' Prentice Hall, New Jersey (2002), pp.197

\bibitem{Jensen1} S. Simic and B. Almohsen, ``Some Generalizations of Jensen's Inequality,'' \href{https://doi.org/10.37256/cm.212021686}{Contemp. Math. \textbf{2}, (2021) 1-14}

\bibitem{Jensen2} F. Hansen and G.K. Pedersen, ``Jensen's Operator Inequality,'' \href{https://doi.org/10.1112/S0024609303002200}{Bull. Lond. Math. Soc. \textbf{35}, (2003) 553-564}

\bibitem{fin1} J. Saavedra, R. Troncoso, J. Zanelli, ``Degenerate Dynamical Systems,'' \href{https://doi.org/10.1063/1.1389088}{J. Math. Phys. \textbf{42}, 4383-4390 (2001)} 
    
\bibitem{fin2} A. L. Ferreira Junior, N. Pinto-Neto, J. Zanelli, ``Dynamical dimensional reduction in multivalued Hamiltonians,'' \href{https://doi.org/10.1103/PhysRevD.105.084064}{Phys. Rev. D \textbf{105}, 084064 (2022)} 
    
\bibitem{fin3} A. L. Ferreira Junior, N. Pinto-Neto, J. Zanelli, ``Inflation and late-time accelerated expansion driven by $k$-essence degenerate dynamics,'' \href{https://doi.org/10.1103/PhysRevD.109.023515}{Phys. Rev. D \textbf{109}, 023515 (2024)} 
    
\bibitem{fin4} F. de Micheli, Jorge Zanelli, ``Quantum Degenerate Systems,'' \href{https://doi.org/10.1063/1.4753996}{J. Math. Phys. \textbf{53}, 102112 (2012)} 
    












\end{thebibliography}
\end{document}